\newif\ifsingle

\newif\ifproofs

\ifsingle
\documentclass[12pt,draftclsnofoot, onecolumn]{IEEEtran}		
\else		
\documentclass[journal]{IEEEtran}
\fi

\ifsingle

\else

\fi 

\usepackage{amssymb}
\usepackage{bm,amsmath,amssymb,amscd,graphicx,amsthm}
\usepackage{algorithm}
\usepackage{algorithmic}
\usepackage{subfigure}
\usepackage{booktabs}
\usepackage{setspace}
\usepackage{microtype}
\usepackage[numbers,sort&compress]{natbib}
\usepackage{color}
\usepackage[bookmarks,colorlinks]{hyperref}

\usepackage{dsfont}
\usepackage{textcomp}
\usepackage{tabu} 
\usepackage{color}

\usepackage{acronym}
\usepackage{textcomp}

\usepackage{endnotes}
\usepackage{comment}
\makeatletter
\def\enoteheading{\section*{\notesname
		\@mkboth{\MakeUppercase{\notesname}}{\MakeUppercase{\notesname}}}%
	\mbox{}\par\vskip-2.3\baselineskip\noindent\rule{.5\textwidth}{0.4pt}\par\vskip\baselineskip}
\makeatother

\newcommand{\Verified}[1]{\footnote{\textcolor{blue}{#1}}}

\newcommand{\edit}[1]{{\color{black} #1}}
	
\begin{document}
	
\acrodef{doa}[DOA]{direction-of-arrival}

\acrodef{mhr}[MHR]{multidimensional harmonic retrieval}	
\acrodef{ista}[ISTA]{iterative shrinkage thresholding algorithm}	
\acrodef{fista}[FISTA]{fast \ac{ista}}
\acrodef{lista}[LISTA]{learned \ac{ista}}	
\acrodef{amp}[AMP]{approximate message passing}
\acrodef{dnn}[DNN]{deep neural network}
\acrodef{admm}[ADMM]{alternating direction method of multipliers}
\acrodef{fmcw}[FMCW]{frequency modulated continuous wave}
\acrodef{fft}[FFT]{fast Fourier transform}
\acrodef{dft}[DFT]{discrete Fourier transform}
\acrodef{idft}[IDFT]{inverse \ac{dft}}
\acrodef{snr}[SNR]{signal-to-noise ratio}
\acrodef{cs}[CS]{compressed sensing}
\acrodef{1d}[1D]{one-dimensional}
\acrodef{2d}[2D]{two-dimensional}
\acrodef{pd}[$P$-D]{$P$-dimensional}

\acrodef{dof}[DoF]{degrees of freedom}

\acrodef{nmse}[NMSE]{normalized mean squared error}
\acrodef{rsfr}[RSFR]{randomized stepped frequency radar}

\title{Structured LISTA for Multidimensional Harmonic Retrieval}

\if false

\author{
	\IEEEauthorblockN{
		Rong~Fu,
		Yimin Liu\IEEEauthorrefmark{1},
		Tianyao Huang, 
		Xiqin Wang}
	\IEEEauthorblockA{Department of Electronic Engineering, \\
		Tsinghua University,
		Beijing, China\\
		Email: fu-r16@mails.tsinghua.edu.cn, \{yiminliu, wangxq\_ee\}@tsinghua.edu.cn} 
	
}
\fi

\author{Rong~Fu, 
 Yimin~Liu$^*$, 
 Tianyao~Huang,
 and Yonina~C.~Eldar
\thanks{
This work is supported by the National Natural Science Foundation of China (Grant No. 61801258). Part of this paper was presented in part at the IEEE Radar Conference (RadarConf), Boston, MA, USA, April 2019 \cite{Fu2019}.

R. Fu, Y. Liu and T. Huang are with the Department of Electronic
Engineering, Tsinghua University, Beijing, 100084, China (e-mail:
fu-r16@mails.tsinghua.edu.cn, \{yiminliu, huangtianyao\}@tsinghua.edu.cn). 

Yonina C. Eldar is with the Faculty of Math
and CS, Weizmann Institute of Science, Rehovot 7610001, Israel (e-mail: yonina.eldar@weizmann.ac.il).

}
} 

\maketitle


\begin{abstract}
	Learned iterative shrinkage thresholding algorithm (LISTA)\acused{lista}, which adopts deep learning techniques to learn optimal algorithm parameters from labeled training data, can be successfully applied to small-scale \ac{mhr} problems. 
 However, \ac{lista} computationally demanding for large-scale \ac{mhr} problems because the matrix size of the learned mutual inhibition matrix exhibits quadratic growth with the signal length. These large matrices consume costly memory/computation resources and require a huge amount of labeled data for training, restricting the applicability of the \ac{lista} method. 
	
	 In this paper, we show that the mutual inhibition matrix of a \ac{mhr} problem naturally has a Toeplitz structure, which means that the \ac{dof} of the matrix can be reduced from a quadratic order to a linear order. 
	By exploiting this characteristic, 
	we propose a structured LISTA-Toeplitz network, which imposes a Toeplitz structure restriction on the mutual inhibition matrices and applies linear convolution instead of the matrix-vector multiplication involved in the traditional \ac{lista} network. 
	Both simulation and \edit{field test for air target detection with radar} are carried out to validate the performance of the proposed network. 
	For small-scale \ac{mhr} problems, LISTA-Toeplitz exhibits close or even better recovery accuracy than traditional \ac{lista}, while the former significantly reduces the network complexity and requires much less training data. 
	For large-scale \ac{mhr} problems, where \ac{lista} is difficult to implement due to the huge size of the mutual inhibition matrices, our proposed LISTA-Toeplitz still enjoys desirable recovery performance. 

\end{abstract}


\IEEEpeerreviewmaketitle

\acresetall

\section{Introduction}\label{Section-Introduction}

Multidimensional harmonic retrieval has been extensively studied in the signal processing literature. This problem appears in a wide range of applications such as wireless communication channel estimation \cite{Wiley2005Space, 5621984}, beampattern synthesis \cite{7952786}, \ac{doa} estimation \cite{Balakrishnan2004A,Compbeamf,Nion2010Tensor} and range-Doppler estimation \cite{HuangCSFAR}. 
\edit{
	Standard methods for MHR include \ac{fft} spectral estimation \cite{1161895} (periodogram), Welch's method \cite{1161901} (also called the modified periodogram) and subspace based methods\cite{4960030,5986845}.
	According to the Nyquist criterion, harmonic signals are generally uniformly sampled at or above the Nyquist rate to avoid aliasing of the spectrum. 

}

In various applications, it is desirable to minimize the required number \edit{out of Nyquist samples needed for \ac{mhr} \cite{2foldblockToep}. 
Especially for \ac{mhr} problems with large dimensions, as the size of measurements is increasingly associated with the dimension, the demand for sample reduction becomes more significant.
Take \ac{doa} estimation with antenna arrays as an example, where each active antenna transmits and receives signals reflected from one or more moving targets to estimate the direction angles.
The smaller amount of samples implies fewer active antennas, lowering the hardware cost \cite{Tan2014a}. 
}
	
Sparse recovery or \ac{cs} \cite{Eldar2012} has been suggested to reduce the measurement data size, when the sinusoids are spectrally sparse, namely when there is a small number of harmonics. 
Particularly, 
consider a harmonic retrieval problem, where a time-domain signal, consisting of $K$ distinct complex sinusoids, has $M$ Nyquist samples.
\edit{
	When it is assumed that the sinusoid frequencies lie precisely on a set of discrete grids, the signal of interest can be sparsely represented by a discrete basis. 
\ac{cs} suggests that the sparse signal can be recovered from a random subset with a reduced size of $N = O(K \log M)$ out of the $M$ Nyquist samples \cite{2foldblockToep}. 
}
The recovery is completed by finding a sparse representation of the time-domain signal over a discrete dictionary matrix $\bm \Phi \in \mathbb{C}^{N \times M}$, \edit{of which a grid corresponds to a predefined grid}. 
To be specific, \ac{cs} formulates \ac{mhr} as a linear decoding problem
\begin{eqnarray} \label{eq:system}
\bm y =\bm \Phi \bm x + \bm w,
\end{eqnarray}
where $\bm y \in {\mathbb{C}^N}$ is the obtained sub-Nyquist samples, $\bm x \in {\mathbb{C}^M}$ denotes the sparse spectral representation of the unknown sinusoids, and $ \bm w \in {\mathbb{C}^N}$ is the additive noise vector. 

A myriad of methods including greedy algorithms and optimization-based approaches 
\cite{Fang2011Greedy, antonello2018proximal, Draganic2017On,Beck2009A,Combettes2006Signal} have been developed to solve such a \ac{cs} problem. Among these algorithms, a key concern for applications is real-time realization. 
While greedy algorithms may lead to a non-optimal solution, optimization-based approaches yield near-optimal estimates in both theory and practice 
\cite{Draganic2017On}. 
One of the well-known $\ell_1$-norm regularization techniques is \ac{ista} \cite{Beck2009A}, which has been proven to have desirable global rate of convergence to the optimum solution \cite{Combettes2006Signal}. 
An accelerated method, namely \ac{fista}, can speed up the rate of convergence 
by adding a momentum term \cite{Beck2009A}. 
The success of \ac{ista} and its variants hinge upon the computation of the proximal operator, which is efficient in a wide range of applications \cite{antonello2018proximal}. 
However, it takes many iterations for \ac{ista} or \ac{fista} to reach a sparse representation, 
which inevitably gives rise to high computational cost thus restricting the application of \ac{cs}.

The approximation ability of deep learning motivates to consider the possibilities of recovering sparse signals at a small computation cost through a neural network. 
By unfolding the iterations of \ac{ista}, \ac{lista} was proposed by Gregor and LeCun \cite{Gregor2010Learning}, which has been demonstrated to be superior to \ac{ista} in convergence speed in both theoretical analysis and empirical results \cite{AMP-Inspired,OnsagerLAMP,Fu2019}. 
{
Following this idea, many researchers \cite{ReviewonUnrolling,OnsagerLAMP,AMP-Inspired,NIPS2016_6406} unfold other iterative algorithms such as \ac{amp} and \ac{admm} algorithms, and incorporate them into deep networks, which have already shown their efficiency in many sparse recovery problems.
It is common for these reconstruction networks to learn some intended network variables for the model information in the dictionary matrix $ \bm \Phi$. 
However, in some cases like large-scale \ac{cs} problems, the dictionary $ \bm \Phi$ has many columns so that there will be a huge amount of network variables to learn. For example, the size of one weight matrix (termed mutual inhibition matrix), which depends on the Gram matrix $\bm \Phi^H \bm \Phi$, to learn in \ac{lista} is up to $M^2$. 
It is very difficult to train a network with high-dimensional variables: Not only will it take much more training time and memory, but also we need to provide a larger training dataset to avoid overfitting. 
}

\edit{
	To reduce the number of trainable parameters, many convolutional extensions based on \ac{lista} were proposed for the applications of optical image super-resolution, denoising or inpainting \cite{Sreter2017Learned, 8808885, LSPARCOM}. 
	These networks exploit the fact that  dictionary matrices in these applications are Toeplitz matrices or concatenations of Toeplitz matrices, which allows one imposing the Toeplitz structure constraint on the learned matrices in \ac{lista} and replacing the matrix-multiplication operations in \ac{lista} with convolutions. Since the \ac{dof} of a $M \times M$ Toeplitz matrix are $O(M)$, much less than the counterpart of a general matrix with the same dimension, such convolutional networks reduce the number of variables to learn significantly. 
	Referring to MHR problems where the dictionaries have a Fourier instead of Toeplitz structure, these convolutional extensions are not directly applicable. When we implement a heuristic convolutional network termed ConvLISTA by directly imposing convolutional prior on the dictionary like \cite{Sreter2017Learned, 8808885, LSPARCOM}, the recovery performance of such a 
	convolutional extension turns to be bad. 

	
	In this paper, we put forward a structured network called LISTA-Toeplitz for \ac{mhr} problems, in order to address the high-dimensional setting. 
	In \ac{mhr}, the Gram matrix $ {\bm \Phi ^H \bm \Phi } $ (rather than the matrix $\bm \Phi$ itself) is a Toeplitz matrix. 
	We recall that this Gram matrix composes the mutual inhibition matrix in \ac{lista}, whose entries are learned from training data.  
	Tailored for \ac{mhr} problems, we shrink the traditional \ac{lista} by imposing Toeplitz structure constraint on the mutual inhibition matrix and replacing the corresponding matrix-multiplication operations with convolution filters, while  the rest components in \ac{lista} are inherited. This  constructs our LISTA-Toeplitz, and is different from previous approaches (e.g., convLISTA) where the Toeplitz structure and convolution operation are related to the dictionary itself. The proposed network  yields a fast and accurate reconstruction on both synthetic and real data, and outperforms ConvLISTA.

}

Main contributions of this paper are summarized as follows:
\begin{enumerate}
 \item 
 We exploit the Toeplitz structure in \ac{1d} harmonic retrieve problems, as well as the doubly block-Toeplitz structure in \ac{2d} harmonic retrieval problems, and thus 
 propose a new structured network called LISTA-Toeplitz, which imposes a Toeplitz structure restriction on the learned mutual inhibition matrices in \ac{lista}. LISTA-Toeplitz significantly shrinks the number of network variables, simplifying the complexity of the network and making the network more trainable. 
 
 \item 
 We use linear convolution to calculate the multiplication by the Toeplitz matrices, which further facilitates the realization of networks. 
 Using convolution instead of matrix-vector multiplication relieves the storage burden, while some fast algorithms for linear convolution reduce computation complexity. 
 Besides, some off-the-shelf neural network toolboxes provide efficient convolution operators, easing the construction of such a structured network.

 \item 
 Both simulated and real data validate the effectiveness of our proposed network and demonstrate its better recovery performance over the traditional LISTA and its previous convolutional extension, i.e., ConvLISTA network. 
 
\end{enumerate}

The rest of this paper is organized as follows.
The signal model in MHR, the conventional \ac{cs} solutions, and our motivations to use \ac{lista} and \ac{lista}-Toeplitz, are introduced in Section \ref{sec:basicsolution}. 
In Section \ref{sec:design} we explore the Toeplitz structure in harmonic retrieval problems and develop the corresponding LISTA-Toeplitz network. 
To show the effectiveness of the proposed networks, in Section~\ref{sec:application} we apply our network to \ac{1d} and \ac{2d} harmonic retrieval problems with both synthetic and real data.
Section~\ref{sec:conclusion} concludes the paper.

\textit{Notation}: 
The symbol $\mathbb{C}$ represents the set of complex numbers. Correspondingly, $\mathbb{C} ^{M}$ and $\mathbb{C} ^{M \times N}$ are the sets of the $M$-dimensional ($M$-D) vectors and $M \times N$ matrices of complex numbers, respectively.
The subscripts $[\cdot]_{i}$ and $[\cdot]_{i,k}$ are used for the $i$-th entry of a vector and the $i$-th row, $k$-th column entry of a matrix. We let $[\cdot]$ and $\{\cdot\}$ denote a vector/matrix and a set, respectively. 
We use a set in subscript to construct a vector/matrix or set, e.g., for a set $\mathcal{N}:=\{0,1,\dots,N-1\}$ and vectors $\bm x_n \in \mathbb{C}^{M}$, $n \in \mathcal{N}$, $[\bm x_n]_{n \in \mathcal{N}}$ and $\{\bm x_n\}_{n \in \mathcal{N}}$ representing the matrix $[\bm x_0, \bm x_1,\dots,\bm x_{N-1}] \in \mathbb{C}^{M \times N}$ and the set $\{\bm x_0, \bm x_1,\dots,\bm x_{N-1}\}$, respectively.
The transpose and Hermitian transpose are written as the superscripts $(\cdot)^T$ and $(\cdot)^H$, respectively.
For a vector, $\| \cdot \|_0$ and $\| \cdot \|_q$ denote the $\ell_0$ ``norm'' and $\ell_q$ norm, $q \ge 1$, respectively.
Operators $*$, $\circ$ and $\otimes$ represent linear convolution, element-wise multiplication and Kronecker product \cite{Schcke2013OnTK}, respectively.
$
$

\section{Motivation and Problem Formulation} \label{sec:basicsolution}
Here, we introduce the motivation for proposing the \ac{lista}-Toeplitz algorithm, i.e., accelerating the \ac{ista}-based methods for \ac{mhr} problems. 
To this aim, we first review preliminaries on the typical \ac{ista} approach and its variations, including \ac{fista} and \ac{lista}, in Subsection \ref{subsec:LISTA}. Then in Subsection \ref{subsec:MHR}, we formulate the \ac{mhr} signal model and show that the inherent Toeplitz structure in \ac{mhr} model can be exploited to improve existing iterative algorithms, yielding the \ac{lista}-Toeplitz algorithm. 
The flow of \ac{lista}-Toeplitz will be detailed in Section \ref{sec:design}.

\subsection{Preliminaries} \label{subsec:LISTA}

In this subsection, we review the procedures of \ac{ista}, \ac{fista} and \ac{lista}, which are designed to solve the \ac{cs} problem \eqref{eq:system}. 

In the framework of \ac{cs}, the dictionary matrix $\bm \Phi$ in \eqref{eq:system} is usually under-determined. 
\ac{ista} methods harness the sparse prior to estimate $\bm x$ by using regularized regression
\begin{eqnarray} \label{eq:regression}
\mathop { \min }\limits_{\bm x} f\left( {\bm \Phi \bm x, \bm y} \right) + \lambda {\left\| \bm x \right\|_1},
\end{eqnarray}
where $f\left( {\bm \Phi \bm x, \bm y} \right) = \frac{1}{2}\left\| {{\bm y - \bm \Phi \bm x}} \right\|_2^2$, measures the error, and $\lambda$ is a regularization parameter controlling the sparsity penalty characterized with $\ell_1$ norm.


The standard \ac{ista} iteratively performs Lipschitz gradient descent with respect to the cost function in \eqref{eq:regression} \cite{Beck2009A,Wu2020Sparse}. 
Specifically, the sparse solution in the $(t+1)$-th iteration, denoted $\bm x^{(t+1)}$, is pursued by the following recursion:
\begin{eqnarray}\label{eq:ISTA}
\begin{aligned}
\bm x^{(t+1)} = {\mathcal{S}_{\frac{\lambda }{L}}}\left( {\frac{1}{L}{\bm \Phi ^H}\bm y + \left( {\bm I - \frac{1}{L}{\bm \Phi ^H}\bm \Phi } \right){\bm x^{(t)}}} \right),\\
\end{aligned}
\end{eqnarray}
where $\lambda$ is the regularization parameter in \eqref{eq:regression}, $L$ is the Lipschitz constant, given by $L = {\lambda _{\max }}( {{\bm \Phi ^H}\bm \Phi } )$, and $\lambda _{\max }(\cdot)$ represents the maximum eigenvalue of a Hermitian matrix. The element-wise soft-threshold operator $\mathcal{S}$ is defined as
\begin{equation}\label{eq:Soperator}
\left[ {\mathcal{S}_{\theta} }\left( \bm{u} \right) \right]_i= \text{sign} \left( {{[\bm{u}]_i}} \right)\left( {\left| {{[\bm{u}]_i}} \right| - \theta} \right)_ + ,
\end{equation}
where sign$ (\cdot)$ returns the sign of a scalar, $ (\cdot)_+ $ means $ \max(\cdot,0)$, 
and $ \theta $ is the threshold. 

\ac{ista} shows its accuracy in recovering sparse signals. However, it may take thousands of iterations for convergence \cite{Draganic2017On}, which motivates development of many variants of \ac{ista} with a higher convergence rate, including the well-known \ac{fista} and \ac{lista} as introduced below.
{
\ac{fista} described in \cite{Beck2009A} can be viewed as a Nesterov’s accelerated version of \ac{ista}, 
\if false
Slightly different from \eqref{eq:ISTA}, \ac{fista}'s basic process in one iteration is 
\begin{eqnarray}\label{eq:FISTA}
\begin{aligned}
{\bm{x}^{(t + 1)}} = \text{prox}_{{\lambda }{\gamma}{{\left\| \bm x \right\|}_1}}&\left( \bm x^{(t)} - \gamma \frac{{\partial f(\bm \Phi \bm{x}^{(t)},\bm y)}}{\partial \bm x} \right.\\
&\quad \left.+ \frac{t-2}{t+1} \left(\bm{x}^{(t)} - \bm{x}^{(t - 1)}\right)\right).
\end{aligned}
\end{eqnarray}
Using the above dedicated step size, \ac{fista} 
\fi
which takes roughly one order-of-magnitude fewer iterations than \ac{ista} \cite{Draganic2017On}. 
}

\begin{figure*} [t]
	\centering
	\includegraphics[width=0.75\textwidth]{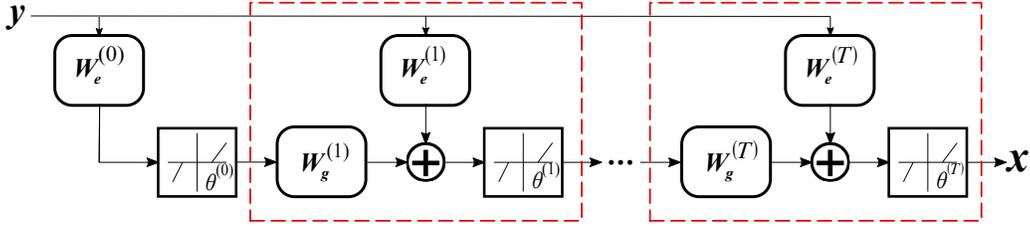}
	\caption{Block diagrams of LISTA network, in which the red broken-line boxes indicate the whole process of one \ac{ista} iteration.}
	\label{fig:frameLISTA}
\end{figure*}
To further accelerate \ac{ista}, Gregor and LeCun proposed a neural network containing only several layers, named \ac{lista} \cite{Gregor2010Learning}. Each layer is an unfolded version of the \ac{ista} iteration \eqref{eq:ISTA} that can be rewritten as
\begin{eqnarray}\label{eq:LISTA}
\bm{x}^{(t + 1)} = \mathcal{S}_{{\theta}^{(t)}}\left( {\bm W_e^{(t)}}\bm y + {\bm W_g^{(t)}}{ \bm{x}^{(t)} } \right),
\end{eqnarray}
where the terms ${\lambda /L}$, ${\frac{1}{L}{\bm \Phi ^H}\bm y}$ and ${\left( {\bm I - \frac{1}{L}{\bm \Phi ^H}\bm \Phi } \right)}$ in \eqref{eq:ISTA} are replaced by ${\theta}^{(t)}$, ${\bm W_e^{(t)}}\in {\mathbb{C}^{M \times N}}$ and ${\bm W_g^{(t)}}\in {\mathbb{C}^{M \times M}}$, respectively. 
The matrices ${\bm W_e^{(t)}} $ and ${\bm W_g^{(t)}} $ are named the \emph{filter matrix} and the \emph{mutual inhibition matrix}, respectively. 
Fig.~\ref{fig:frameLISTA} illustrates an LISTA network structure with $T$-layer. 

Opposed to \ac{ista} and \ac{fista}, where parameters in each iteration are identical and are calculated analytically or set manually, \ac{lista} treats the tuple $(\bm W_e^{(t)},\bm W_g^{(t)},\theta^{(t)} )$ in each layer, $t = 0,1,\dots,T-1$, as variables to learn from some predefined training data using the back-propagation algorithm.
\edit{The network performs better when making the values of parameters differ in each layer. We may omit the superscript ${(t)}$ in the following notation for simplicity.}
Numerical results show that \ac{lista} can achieve virtually the same accuracy as the original \ac{ista} using nearly two orders of magnitude fewer iterations \cite{AMP-Inspired,8462136}.

Nevertheless, a challenge in \ac{lista} is that there are many variables to learn. 
A large neural network leads to high computational burden. Besides, it requires careful tuning of hyper-parameters such as learning rates and initialization values to avoid training problems such as over-fitting \cite{Overfitting} and gradient vanishing \cite{DBLP:journals/corr/IoffeS15}. 
In \ac{cs} problems with large-scale sparse signal $\bm x \in \mathbb{C}^M$, the mutual inhibition matrix ${\bm W_g}$ of size $M \times M$, 
\edit{which is much larger than $\bm W_e$ thus }
takes a dominant role among all the variables, especially when allowing ${\bm W_g}$ varying across the layers. 
It motivates us to tailor the network to fit a specific problem so that we can impose restrictions on ${\bm W_g}$ in order to reduce the computational burden.
For some certain \ac{cs} problems like \ac{mhr}, the mutual inhibition matrix naturally has a Toeplitz structure, which can be exploited to reduce the dimension of neural networks, as will be detailed in the subsequent subsection.

\subsection{Toeplitz Structure in MHR Problems} \label{subsec:MHR}

Here, we introduce the signal model of \ac{mhr} problems, associated with many practical applications including \ac{1d}, \ac{2d} \ac{doa} estimation and range-Doppler recovery in radar systems. Then, we reveal that \ac{mhr} has specific characteristics of the dictionary matrices, which can help us reduce the dimensions of the learned variables in \ac{lista}.
The Toeplitz structures of \ac{1d} and \ac{2d} harmonic retrieval problems are discussed in Subsections \ref{subsec:1dMHR} and \ref{subsec:2dMHR}, respectively.
\subsubsection{Toeplitz structure of 1D harmonic retrieval} \label{subsec:1dMHR}

{
First, we consider a \ac{1d} harmonic retrieval problem, following the signal model presented in \cite{2foldblockToep}.
Assume that there are $K$ distinct complex-value sinusoids, with their frequencies denoted by $ f_k$, $k = 1,2,\dots,K$. 
The observation taken at $\bm i$-th time instant can be reformed into a superposition of these $K$ complex-value sinusoids, given by
\begin{equation}
y_{i} = \sum \limits_{k = 1}^{K} a_k e^{j2\pi f_k i},
\label{eq:K_sines_1D}
\end{equation}
where $a_k$ denotes the complex amplitude of the $k$-th sinusoid. 

We uniformly discretize frequencies into $M_1$ grids and assume that the $K$ sinusoids are on the grids.
Thus we recast \eqref{eq:K_sines_1D} in matrix form as below
\begin{eqnarray} \label{eq:MD_dict}
\bm y^{\star} = \bm \Psi \bm x,
\end{eqnarray}
where $\bm x \in \mathbb{C}^M$ contains only $K$ nonzero elements, corresponding to the complex des of the $K$ sinusoids, and $\bm \Psi$ is the $M_1 \times M_1$ discrete Fourier matrix $ \bm F_{M_1}$ whose $(i, m)$-th entry is defined as
\begin{equation} \label{eq:Fourier matrix}
[\bm \Psi]_{i,m} = [\bm F_{M_1}]_{i,m} = e^{\mathrm{j} 2 \pi \frac{ m}{M_1} i }, i_p,m\in \mathcal{M}_1.
\end{equation}

Furthermore, we consider a compressive measurement, where only $N$ entries of $\bm y^{\star}$ are observed with $N \ll M_1$. 
To store the indices of selected entries from $\bm y^{\star}$, we define a subset $\Omega$ of cardinality $N$ randomly chosen from the set $\mathcal{M}_1$.
Then we use a row-subsampled matrix $\bm R$ to selects $N$ rows of $\bm \Psi$ corresponding to the elements in $\Omega$, i.e., $\left[ \bm R \right]_{n,m} = 1$,where $m$ is the $n$-th element of $\Omega$ while other entries in the $n$-th row being zeros. 
Thus, the sub-sampled observations are denoted by $\bm y \in \mathbb{C}^{N}$, i.e.,
\begin{equation}
\bm y = \bm R \bm y^{\star}
= \bm R \bm \Psi \bm x,
\label{eq:partial_model}
\end{equation}
Here, we use $ {\bm \Phi} = \bm R \bm \Psi$ to represent the new dictionary matrix, consisting of sub-sampled $N$ rows of the full dictionary $\bm \Psi$.

The Gram matrix of this dictionary matrix is $ \bm \Phi ^H \bm \Phi = {\bm F_{M_1} ^H \bm R ^H }{\bm R \bm F_{M_1}}$,
which can be formulated as 
\begin{eqnarray}
\bm \Phi ^H \bm \Phi = \sum\limits_{m = 0}^{M-1} {\mathds{1}_\Omega(m) \bm \phi_m \bm \phi_m^H },
\end{eqnarray}
where $\mathds{1}_\Omega(m)$ is an indicator function which equals 1 when $m \in \Omega$ and zero otherwise, and $\bm \phi_m$ is the $m$-th column of the Fourier matrix $\bm F_{M_1}$. It can be verified that $\bm \phi_m \bm \phi_m^H $ is a Hermitian Toeplitz matrix.
As a consequence, the Gram matrix above can be viewed as a sum of $N$ Hermitian Toeplitz matrices, which is still a Hermitian Toeplitz matrix. 
Note that for a Hermitian Toeplitz matrix, the \ac{dof} is $M$. If we disregard the Hermitian symmetry, the \ac{dof} is slightly larger, $2M-1$. In both cases, the \ac{dof} are much less than $M^2$, that of a general matrix without such structure, which can be exploited to reduce the complexity of networks in \ac{lista}.

}

\subsubsection{Doubly-block Toeplitz structure of 2D harmonic retrieval} \label{subsec:2dMHR}

For the 2D harmonic retrieval, the corresponding dictionary matrix is
$\bm \Phi = \bm R \left( \bm F_{M_1} \otimes \bm F_{M_2} \right) $. 

{	
	The Gram matrix can be formulated as
	\begin{eqnarray*}
		\begin{aligned}
			\bm \Phi^H &\bm \Phi 
			= \left( \bm F_{M_1} \otimes \bm F_{M_2} \right)^H \bm R^H \bm R \left( \bm F_{M_1} \otimes \bm F_{M_2} \right)\\	
			& \stackrel{(a)}{=} 
			\sum\limits_{m_2 = 0}^{M_2-1} { 
				\sum\limits_{m_1 = 0}^{M_1-1} {
					\mathds{1}_\Omega \{(m_1,m_2)\} 
					\left ( \bm \phi_{m_1} \otimes \bm \psi_{m_2} \right ) 
					\left ( \bm \phi_{m_1}^H \otimes \bm \psi_{m_2}^H \right) } }\\
			& \stackrel{(b)}{=} 
			\sum\limits_{m_2 = 0}^{M_2-1} { 
				\sum\limits_{m_1 = 0}^{M_1-1} {
					\mathds{1}_\Omega \{(m_1,m_2)\} 
					\left ( \bm \phi_{m_1} \bm \phi_{m_1}^H \right ) \otimes
					\left ( \bm \psi_{m_2} \bm \psi_{m_2}^H \right) } },
		\end{aligned}
	\end{eqnarray*}
	where $\bm \phi_{m_1}$ and $ \bm \psi_{m_2}$ denote the $m_1$-th, $m_2$-th column of the Fourier matrix $\bm F_{M_1}$ and $\bm F_{M_2}$, respectively. 
	The equation (a) is a consequence of the fact that taking the complex conjugate or transpose before carrying out
	the Kronecker product yields the same result as doing so afterward, and (b) stems directly from the mixed product property of Kronecker product \cite{Schcke2013OnTK}.
	
	Similarly, both $\bm A := \bm \phi_{m_1} \bm \phi_{m_1}^H \in \mathbb{C}^{M_1 \times M_1}$ and $\bm B := \bm \psi_{m_2} \bm \psi_{m_2}^H \in \mathbb{C}^{M_2 \times M_2}$ are Hermitian Toeplitz matrices.
	Thus, the Toeplitz matrix $\bm A$ can be represented by a $(2M_1-1)$-D vector, 
	denoted by $[a_l]_{l \in \mathcal{M}_1^{\ast} }$, 
	where 
	$\mathcal{M}_1^{\ast}:=\{ -M_1+1,-M_1+2,\cdots, M_1-1\}$ and we use the asterisk notation in the superscript to distinguish from the set $\mathcal{M}_p$. Particularly, the $(i,j)$-th element of $\bm A$ can be denoted as $[\bm A]_{i,j} = a_{i-j}$.
	According to the definition of Kronecker product, we have
	\begin{equation*}
	\bm A \otimes \bm B = 
	\left[ {\begin{array}{*{20}{c}}
		{\bm C_{0}}&{\bm C_{-1}}& \cdots &{\bm C_{ -{M_1}+ 1}}\\
		{\bm C_{1}}&{\bm C_{0}}& \ddots & \vdots \\
		\vdots & \ddots & \ddots &{\bm C_{ - 1}}\\
		{\bm C_{{M_1} - 1}}& \cdots &{\bm C_{1}}&{\bm C_{0}}
		\end{array}} \right],
	\end{equation*}
	where $ \bm C_i = { a_i} \bm B$, $i \in \mathcal{M}_1^{\ast} $.
	
	The matrix $ \bm A \otimes \bm B $ has a so-called doubly-block Toeplitz structure \cite{blockToep1,blockToep2}, i.e., each block matrix of which is itself a Toeplitz matrix and also repeated down the diagonals of the whole matrix.
}
In the literature, such matrix is also called two-fold
block Toeplitz matrix \cite{2foldblockToep} or Toeplitz-block-Toeplitz matrix \cite{Toep-Block-Toep}. After the sum operation over $m_1$ and $m_2$, the doubly-block Toeplitz structure is preserved, implying that the Gram matrix in the \ac{2d} \ac{mhr} problem is also a doubly-block Toeplitz. Observing the structure of $ \bm \Phi^H \bm \Phi $, we find that it is constructed by two Hermitian Toeplitz matrices, of which the \acp{dof} are $ 2M_1-1 $ and $ 2M_2-1 $, respectively. Consequently, the \ac{dof} of the Gram matrix $ \bm \Phi^H \bm \Phi $ is $ (2M_1-1)(2M_2-1) $, much less than $ M^2 = M_1^2 M_2^2 $, the number of elements in $ \bm \Phi^H \bm \Phi$
, indicating the possibility to compress the networks in the original \ac{lista}.



{
	Previous discussions presented in Subsections \ref{subsec:1dMHR} and \ref{subsec:2dMHR} reveal that the Gram matrix $ \bm \Phi^H \bm \Phi $ possesses Toeplitz or Toeplitz related structure. Hence, the mutual inhibition matrix $\bm W_g$ to learn, which corresponds to the ${\left( {\bm I - \frac{1}{L}{\bm \Phi ^H}\bm \Phi } \right)}$, is also Toeplitz structured and thus compressible. 
	Inspired by this phenomenon in \ac{mhr} problems, we propose a heuristically structured network called LISTA-Toeplitz
	by imposing a Toeplitz structure on ${\bm W_g}$ in \ac{lista}. 
	Applying such a structure-imposing approach for neural network will benefit performance by compression and model reduction in space complexity and gain high sample efficiency in training process.
	In the following Section \ref{sec:design}, we will explain how to build the LISTA-Toeplitz network. 
}

\section{LISTA-Toeplitz Network Design} \label{sec:design}

As shown in Section \ref{subsec:MHR}, \ac{1d}/\ac{2d} harmonic retrieval problems naturally possess a Toeplitz/doubly-block Toeplitz structure on the Gram matrix $ \bm \Phi^H \bm \Phi $. In this section, we will impose the Toeplitz structure restriction on the corresponding variables of the \ac{lista} network, yielding the design of \ac{lista}-Toeplitz network. The motivations of using the Toeplitz structure in building network architecture are multi-fold: 
1) Neural networks with structured weight matrices will obtain model order reduction in the number of network variables so that it can be potentially applied to large-scale sparse recovery problems; 
2) Taking advantage of the model information, it performs much more efficiently to train such a model-based network and easier to accomplish the desired result for specific problems, especially when the amount of training data is limited. 
The rest of this section will be devoted to detailed introductions of \ac{lista}-Toeplitz networks for \ac{1d} and \ac{2d} harmonic retrieval problems, shown in Subsections \ref{subsec:1D} and \ref{subsec:2D}, respectively.

\subsection{1D LISTA-Toeplitz Network} \label{subsec:1D}
Here, we consider \ac{1d} harmonic retrieval problems and design the \ac{lista}-Toeplitz network by exploiting the Toeplitz structure in the mutual inhibition matrix ${\bm W_g} \in \mathbb{C}^{M \times M}$, where $M$ is the number of 1D grid points. 

Since a $M$ dimensional Toeplitz matrix can be represented by a $ 2M-1 $ dimensional vector, we denote such vector by $\bm h: = [ h_{m}]_{m \in \mathcal{M}^{\ast}}^T\in\mathbb{C}^{2M-1}$. Consequently, the matrix $ \bm W_g $ is expressed by ${[\bm W_g]_{i,k}} = h_{i-k}$, $i,k \in \mathcal{M}$, shown in the following equation
\begin{eqnarray}
\label{eq:Wgh}
\bm W_g = 
\left[ {\begin{array}{*{20}{l}}
	{{ h_0}}&{{ h_{ - 1}}}&{{ h_{ - 2}}}& \cdots &{{ h_{ - M+ 1}}}\\
	{{ h_1}}&{{ h_0}}&{{ h_{ - 1}}}& \ddots & \vdots \\
	{{ h_2}}& \ddots & \ddots & \ddots &{{ h_{ - 2}}}\\
	\vdots & \cdots & \ddots & \ddots &{{ h_{ - 1}}}\\
	{{h_{M- 1}}}& \cdots &{{ h_2}}&{{ h_1}}&{{ h_0}}
	\end{array}} \right].
\end{eqnarray}
Here, we disregard the Hermitian structure of $ \bm W_g $, to achieve a compromise between the expressive power and compactness of the network architecture.

Based on the Toeplitz structure of $ \bm W_g $, the multiplication operation $ \bm W_g \bm x $ in \eqref{eq:LISTA} can be expressed by a linear convolution between two vectors and realized by some off-the-shelf toolboxes 
\cite{MLframeworks,TensorFlowGuide}. To see this, note that the linear convolution between $\bm h $ and $\bm x \in \mathbb{C}^M $, denoted $ \bm h * \bm x $, yields a $ M $ dimensional vector with entries given by 
\begin{equation}
\label{eq:conv_equ}
{\left[ {\bm h * \bm x} \right]_i} = \sum\limits_{k = 0}^{M-1} {{ h_{i - k}}{[\bm x]_k}},i \in \mathcal{M}.
\end{equation}
The right hand side in \eqref{eq:conv_equ} also equals to $ \left[\bm W_g \bm x\right]_i $, according to definition of $ \bm W_g $ in \eqref{eq:Wgh} and the rule of matrix multiplication. Hence, we have
\begin{equation}
\label{eq:multiplicationandconv}
{\bm W_g}\bm x = \bm h * \bm x.
\end{equation}

\begin{figure*} [tb]
	\centering
	\includegraphics[width=0.75\textwidth]{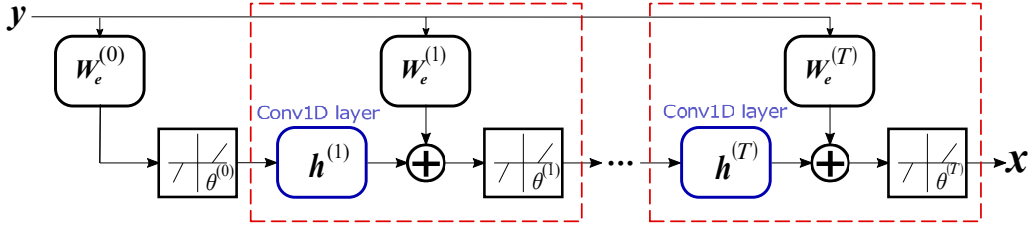}
	\caption{The Block diagram of \ac{1d} \ac{lista}-Toeplitz network. The red broken-line boxes indicate each layer of \ac{lista}-Toeplitz network, corresponding to the whole process of an \ac{ista} iteration. Blue boxes represent convolutional layers \eqref{eq:multiplicationandconv}, which highlight the modification of \ac{lista}-Toeplitz over the standard \ac{lista}. }
	\label{fig:1dLISTA-frame}
\end{figure*}

Instituting \eqref{eq:multiplicationandconv} into \eqref{eq:LISTA} implies a proximal mapping using linear convolution operation, given by
\begin{eqnarray}\label{eq:LISTA_Toep_1D}
{\bm x^{(t+1)}} = {\mathcal{S}_{{\theta ^{(t)}}}}\left( {\bm W_e\bm y + {\bm h } * {\bm x^{(t)}}} \right).
\end{eqnarray}
Based on \eqref{eq:LISTA_Toep_1D}, we replace each layer of the standard \ac{lista} with a structured network, which yields the \ac{1d} version of proposed \ac{lista}-Toeplitz. The whole framework the framework of our proposed LISTA-Toeplitz network is given in Fig. ~\ref{fig:1dLISTA-frame}.

Comparing Figs.~\ref{fig:frameLISTA} and~\ref{fig:1dLISTA-frame}, we note that the difference of the conventional \ac{lista} and \ac{lista}-Toeplitz lies in the realization of multiplication $\bm W_g \bm x $. The latter uses a dimension-reduced vector $ \bm h $ to replace the large matrix $ \bm W_g $, and applies linear convolution \eqref{eq:multiplicationandconv}.
The reduction in space complexity and computational complexity by using structured matrices are significant. The advantages of \ac{lista}-Toeplitz are discussed in detail as follows.

{
	1) For the space complexity, the memory demand of the network is reduced, which also contributes to lower the cost and power consumption of the networks \cite{Potok2017A}. 
	While the 1D LISTA needs to update the weight matrix $\bm W_g$ of size $M^2$, our proposed 1D LISTA-Toeplitz network only needs to learn $2M-1$ elements for $\bm W_g$. 
	Thus, the space complexity of our LISTA-Toeplitz network is $O(M)$. This is a significant advantage compared to the original LISTA network which requires $O(M^2)$ parameters. Especially for large-scale harmonic retrieval problems, our proposed LISTA-Toeplitz network decreases memory requirements by a factor of $M$, which greatly relieves the storage burden.

	2) Resorting to linear convolution operation also reduces the computational burden. 
	For computation efficiency, we can use Discrete Fourier Transform (DFT) to speed up the computation process of linear convolution. Particularly, the linear convolution in \eqref{eq:LISTA_Toep_1D} can be computed in the Fourier domain, given by
	\begin{eqnarray}
	{\bm x^{(k+1)}} = {\text{S}_{{\theta ^{(k)}}}}\left( {\bm W_e}\bm y + \mathcal{F}^{-1}\left( \mathcal{F}(\bm h) \circ \mathcal{F}(\bm x^{(k)}) \right) \right),
	\end{eqnarray}
	where $\circ$ denotes element-wise multiplication and $\mathcal{F}$ denotes the DFT operator while $\mathcal{F}^{-1}$ is the inverse DFT (IDFT) operator. DFT and IDFT operators can be efficiently computed with Fast Fourier Transforms \cite{OppenheimDSP,Ye2018}, and the time complexity is $O(M \log M)$ in the term of required times of complex-value multiplications. Therefore, the proposed 1D LISTA-Toeplitz network has time complexity $O(M \log M)$ in each layer, more efficient than the traditional \ac{lista} enjoying the counterpart of $O(M^2)$.

	3) Because of the huge reduction in the network variables, the proposed structured network is more trainable for large-scale problems and performs better on limited training data. 
	It has been commonly considered that the number of training samples need to be more than roughly ten times the number of network variables \cite{ZhuVFR15,VanSampleSize,lei2019how}. 
	Compared with its unstructured counterpart, \ac{lista}-Toeplitz reduces the number of network variables, therefore requires less labeled data for training. Consequently, this takes less training time and makes the process of training neural networks more efficiently. 
	
	To summarize, different from other network compression approaches such as Principal Filter Analysis (PFA) which provides a specific or heuristic compression factor \cite{suau2019network}, the Toeplitz constraint used in the proposed methods enables model order reduction and provides huge computational complexity reduction of large-scale problems: The storage requirement is reduced from $O(M^2)$ to $O(M)$ and the computational complexity can be reduced from $O(M^2)$ to $O(M \log M)$. Table~\ref{table1} compares the time and space complexity of LISTA and 1D LISTA-Toeplitz network.
	\begin{table} 
		\caption{The one-layer complexity of the LISTA and LISTA-Toeplitz networks.} \label{table1}
		\begin{tabu} to 0.5\textwidth{X[7,c]|X[5,b]|X[4,l]} 
			\hline 
			Network &Time &Space \\ 
			\hline 
			LISTA &$O(M^2)$ &$O(M^2)$ \\ 
			LISTA-Toeplitz &$O(M \log M)$ &$O(M)$ \\ 
			\hline 
		\end{tabu} 
	\end{table} 
	These benefits are also applicable to higher dimensional harmonic retrieval problems, though the network architecture is slightly different, as shown in the subsequent subsection.
}

\subsection{2D LISTA-Toeplitz Network} \label{subsec:2D}
In this subsection, we extend the \ac{lista}-Toeplitz network to \ac{2d} harmonic retrieval problems. 
As shown in Subsection~\ref{subsec:2dMHR}, the mutual inhibition matrix $ \bm W_g $ naturally follows a doubly-block Toeplitz structure, which requires a slight change in the network settings of \ac{lista}-Toeplitz. For clarity, we denote such a network by \ac{2d} \ac{lista}-Toeplitz. 
Recall that in the \ac{2d} harmonic retrieval problems, we use $ M_1 $ and $ M_2 $ to denote the cardinality of the grid sets in the first and second dimension, respectively, and $ M = M_1 M_2 $. 
The sparse vector $ \bm x $ has a block structure, expressed as $ \bm x = \left[\bm x_{m_2}^T\right]_{m_2 \in \mathcal{M}_2}^T $, where the sub-vector $ \bm x_{m_2} $ is denoted by $ \bm x_{m_2} = \left[x_{m_1,m_2}\right]_{m_1 \in \mathcal{M}_1}^T \in \mathbb{C}^{M_1} $.

Following the similar procedure for \ac{1d} case, we first use a dimension reduced matrix $\bm H \in \mathbb{C}^{(2M_1 -1)\times (2M_2 -1)}$ to represent $ \bm W_g \in \mathbb{C}^{M_1M_2 \times M_1M_2}$. This is possible, because the \ac{dof} of $ \bm W_g $ is $ (2M_1 -1)\times (2M_2 -1)$ due to its Toeplitz structure. 
For convenience, $ \bm H $ is expressed as $ \bm H :=\left[h_{m_1,m_2}\right]_{m_1 \in \mathcal{M}_1^{\ast}, m_2 \in \mathcal{M}_2^{\ast}} $. Here we also ignore the Hermitian structure for making some compromise on the representational probability of the LISTA-Toeplitz network.

To link between $ \bm H $ and $ \bm W_g $, we first define $ 2M_2-1 $ Toeplitz sub-matrices $ \bm H_m \in \mathbb{C}^{M_1 \times M_1} $ constructed from the $m$-th column of $\bm H$, i.e., $[h_{m_1,m}]_{m_1 \in \mathcal{M}_1}^T$, where $ m \in \mathcal{M}_2 $, given by $\left[ \bm H_m \right]_{i,j} = h_{i-j,m}$.

With these Toeplitz matrices, we construct a doubly-block Toeplitz matrix $\mathcal{T}(\bm H) \in {\mathbb{C}^{M_1M_2 \times M_1M_2}}$ as 
\begin{eqnarray}
\mathcal{T}(\bm H) := \left[ {\begin{array}{*{20}{c}}
	{\bm H_{0}}&{\bm H_{-1}}& \cdots &{\bm H_{ -{M_2}+ 1}}\\
	{\bm H_{1}}&{\bm H_{0}}& \ddots & \vdots \\
	\vdots & \ddots & \ddots &{\bm H_{ - 1}}\\
	{\bm H_{{M_2} - 1}}& \cdots &{\bm H_{1}}&{\bm H_{0}}
	\end{array}} \right].
\end{eqnarray}
We can verify that ${\bm W_g} = \mathcal{T}(\bm H)$ if the elements of $\bm H$ satisfy
\begin{eqnarray}\label{eq:H_and_Wg}
h_{s-i, t-j} = \left[\bm W_g\right]_{s+(t-1) \times M_1,i+(j-1) \times M_1},
\end{eqnarray}
where $i,s \in \mathcal{M}_1$ and $j,t \in \mathcal{M}_2$. 

\begin{figure*} [tb]
	\centering
	\includegraphics[width=0.75\textwidth]{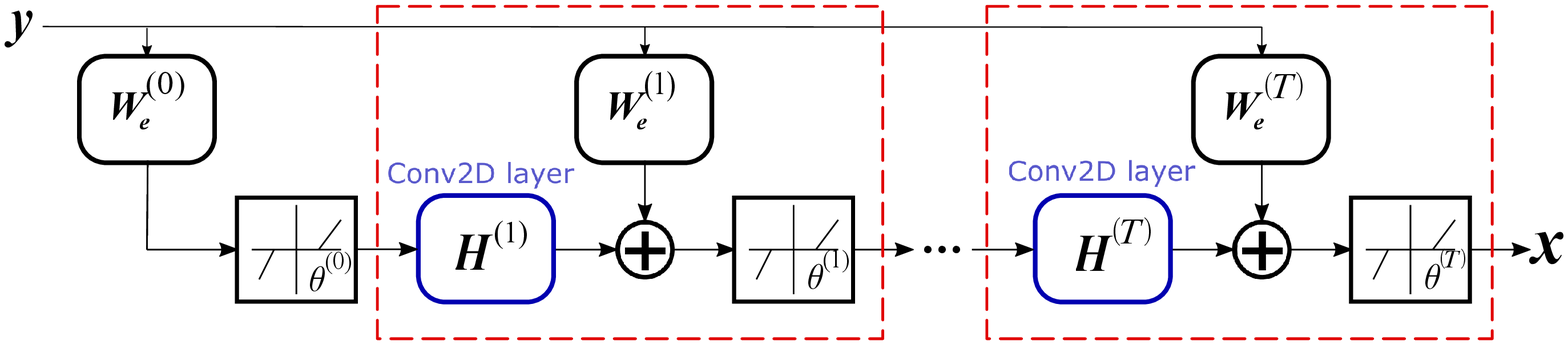}
	\caption{The Block diagram of \ac{2d} \ac{lista}-Toeplitz network. The red broken-line boxes indicate each layer of \ac{lista}-Toeplitz network, corresponding to the whole process of an \ac{ista} iteration. Blue boxes represent linear convolutional layers \eqref{eq:multiplicationandconv}, which highlight the modification of \ac{lista}-Toeplitz over the standard \ac{lista}. }
	\label{fig:2DLISTA-frame}
\end{figure*}

We then prove that the result of $ \bm W_g \bm x $ can be calculated by linear convolution operation as follows.
Since the vector $ \bm x $ has a nested structure, we define $ \bm X :=\left[\bm x_{1} \ \bm x_{2} \ \cdots \ \bm x_{M_2}\right] \in \mathbb{C}^{M_1 \times M_2} $, 
which satisfies that $\bm x = \textrm{vec}(\bm X)$. 
The linear convolution operation between two matrices is defined as
\begin{equation}\label{eq:conv_matrix}
\left[ \bm H *\bm X \right]_{s,t} = 
\sum\limits_{j = 0}^{{M_2-1}} {\sum\limits_{i = 0}^{{M_1-1}} {{h_{s - i,t - j}}{x_{i,j}}} },
\end{equation}
where $s \in \mathcal{M}_1$, $t \in \mathcal{M}_2$. 
According to the definition of matrix multiplication, we have
\begin{equation}\label{eq:multiplication}
\begin{array}{l}
\begin{aligned}

\left[\bm W_g \bm x\right]_{l}
&= \sum\limits_{k = 1}^{{M_1M_2}} {{\left[\bm W_g\right]_{l,k }}\left[\bm x\right]_k} \\
&=
\sum\limits_{j = 1}^{{M_2}} 
{\sum\limits_{i = 1}^{{M_1}} 
	{{\left[\bm W_g\right]_{l,i + j M_1}}{x_{i,j}}} 
} .
\end{aligned}
\end{array}
\end{equation}
{
	Comparing \eqref{eq:conv_matrix} and \eqref{eq:multiplication}, we conclude that $\bm W_g \bm x = \textrm{vec}(\bm H *\bm X)$ by setting $l = s+t M_1$ in the subscript in \eqref{eq:multiplication}.
	Thus, the unfolded \ac{ista} iteration \eqref{eq:LISTA} is rewritten as
	\begin{eqnarray}\label{eq:LISTA_Toep_2D}	
	\begin{aligned}
	{\bm x^{(t+1)}} & = {\mathcal{S}_{{\theta ^{(t)}}}}\left( {\bm W_e\bm y + \textrm{vec}\left(\bm H *\bm X^{(t)}\right)} \right),
	\end{aligned}
	\end{eqnarray}
	where $\bm X^{(t)} \in \mathbb{C}^{M_1 \times M_2}$ is obtained by reshaping $\bm x^{(t)}\in \mathbb{C}^{M_1 M_2}$, i.e., $\bm X^{(t)} = 
	\left[\bm x^{(t)}_{1} \ \bm x^{(t)}_{2} \ \cdots \ \bm x^{(t)}_{M_2}\right] $.
	Based on \eqref{eq:}, we illustrate the \ac{2d} \ac{lista}-Toeplitz network in Fig.~\ref{fig:2DLISTA-frame}. 
}

Same with the \ac{1d} \ac{lista}-Toeplitz, the \ac{2d} version significantly reduces the number of variables in the weight matrix compared with its counterpart of the standard \ac{lista}, from $ O\left( M_1^2 M_2^2 \right) $ to $ O\left( M_1M_2 \right) $.
For a typical case where $ M_{\ast} = M_1 = M_2 $, the reduction in spacial complexity, from $ O\left(M_{\ast}^4\right) $ to $ O\left(M_{\ast}^2\right) $, is more notable than the \ac{1d} case. 
In higher dimension scenarios, with $ M_{\ast} = M_1 = M_2 = \cdots = M_P$, the \ac{lista}-Toeplitz network can be simply extended, and the benefits of exploiting the Toeplitz structure becomes more dominant: The memory demand for storing the variables to learn will be decreased from $ O\left(M_{\ast}^{2P}\right) $ to $ O\left(M_{\ast}^P\right) $. 
As discussed in Subsection \ref{subsec:1D}, the compressed network also contributed to making the network more trainable, lowering the cost and power consumption. 

{
	Note that this convolutional prior is imposed on the Gram matrix of the dictionary matrix rather than the dictionary itself. That is why we replace matrix multiplication with linear convolution just for the mutual inhibition part ${\bm W_g}{ \bm{x}^{(t)} }$. 
	For comparison, we also try to making the linear multiplication by ${\bm W_e}$ to be convolution, although there is no equivalence between ${\bm W_e}\bm y$ and ${\bm h_e}*\bm y$ where ${\bm W_e} \in C^{M \times N}$ and ${\bm h_e} \in C^{(M+N-1) \times 1}$.
	We achieve the architectures of this convolutional network (named ConvLISTA) as follows.	
	\begin{eqnarray}\label{eq:}	
	{\bm x^{(t+1)}} =
	\left\{ 
	{\begin{array}{*{20}c}
		{\mathcal{S}_{{\theta ^{(t)}}}}\left( {{\bm h_e}*\bm y + {\bm h } * {\bm x^{(t)}}} \right), \text{for 1D MHR}\\
		{\mathcal{S}_{{\theta ^{(t)}}}}\left( {{\bm h_e}*\bm y + \textrm{vec}\left(\bm H *\bm X^{(t)}\right)} \right), \text{for 2D MHR}
		\end{array}} 
	\right.
	\end{eqnarray}
	
	In MHR problems, as the dictionary matrix does not have shift-invariant structure, the recovery performance of ConvLISTA, where both linear blocks are implemented by convolutions, is much worse than the LISTA and our \ac{lista}-Toeplitz networks. 
	The numerical results of these networks are shown in next section.

}

\section{Practical Applications and Numerical Results}\label{sec:application}
{
	In this section, we perform numerical experiments involving 1D and 2D harmonic retrieval problems to demonstrate the effectiveness of the proposed LISTA-Toeplitz network in comparison with the original LISTA network, ConvLISTA and conventional iterative algorithms including ISTA and FISTA. 
	As presented in the sequel, we showcase our performance by using both synthetic data and real data\footnote{Codes for reproducing these experiments are available at https://github.com/....}. 
	In Subsections \ref{subsec:1Dsimulation} and \ref{subsec:2Dsimulation}, we generate synthetic data 
	for both 1D and 2D harmonic retrieval problems, respectively. 
	The reconstruction quality of these methods is measured in terms of \ac{nmse} and hit rate.
	Here, \ac{nmse} means the mean squared error of the recovered signal $\hat{\bm x}$ normalized by the power of ground truth $\bm x$, given by
	\begin{eqnarray}\label{eq:NMSE1}
	\mathrm{NMSE} = 
	{\rm E}{ {\left\| {\bm x - \bm{\widehat x}} \right\|_2} }/
	{\left\| \bm x \right\|_2}.
	\end{eqnarray}
	Hit rate is defined as the percentage of the number of correctly recovered components to the total number of nonzero entries. 
	In Subsection \ref{subsec:real data}, we apply our proposed 2D LISTA-Toeplitz network to real radar measurements, which verifies its performance in real data.
}

The tested algorithms are set as follows. Since ISTA and FISTA usually take about 1000 and 100 iterations to converge, respectively, while LISTA, ConvLISTA and our LISTA-Toeplitz only need 10 layers to converge, in the following experiments, we set the numbers of iterations/layers for ISTA, FISTA, LISTA, ConvLISTA and LISTA-Toeplitz as around 1000, 100, 10 and 10, respectively, which will be specified in the following sections. 
For LISTA, ConvLISTA and our LISTA-Toeplitz methods, the performance relies on not only the network structure but also the amount of training data, which will be specified in the following experiments individually. Following the practical criterion in \cite{ZhuVFR15,VanSampleSize,lei2019how}, we generate training data as many as roughly ten times the number of unknown parameters in the network. 
\edit{
	For each training pair, we generate sparse signals $\bm x$ as i.i.d. Bernoulli-Gaussian with the fixed sparsity $K$, where the sparse support follows a Bernoulli distribution and the complex amplitude follows a Gaussian distribution.
	Using a fixed dictionary based on harmonic retrieval model, the observations $\bm y$ are computed according to \eqref{eq:partial_model} under the noise power $\sigma^2$,

}

More details on network settings and training procedures of LISTA-Toeplitz are referred to Appendix, where we introduce special modifications to handle complex-value data in Appendix \ref{subsec:complexextension}, and introduce some training details in Appendix \ref{subsec:training}. For example, we explain how to generate training dataset, how to compute the initial value of network parameters and train our LISTA-Toeplitz network.

\subsection{1D Harmonic Retrieval with Synthetic Data} \label{subsec:1Dsimulation}
{
	To show the effectiveness of the proposed \ac{1d} LISTA-Toeplitz network, we formulate an 1D harmonic retrieval problem in two different cases: noiseless and noisy synthetic data.
}

Here, the measurement vector $\bm y$ is generated using the signal model \eqref{eq:partial_model}, 
where $N=64$ samples are randomly selected from $M = 512$ measurements. The normalized frequency $[0,1)$ is equally divided into $M = 512$ grids, making $\Delta \omega = 2\pi/M$. 
The number of signal components is $K=5$ unless stated specially, and their amplitudes are Gaussian distributed. 
{
	In the training process, the noise power was fixed as $\sigma^2 = 0.4$. 
}
Regarding ConvLISTA and our proposed LISTA-Toeplitz network, we generate 50,000 trials for training and 1000 trials for verification and testing. As a benchmark, we also train a ten-layer LISTA network, which has around $10M^2 \approx 2 \times 10^6 $ parameters to learn. To avoid overfitting, we generate $10^7$ trails for the traditional \ac{lista} network.

\begin{figure}
	\centering
	\includegraphics[width=0.9\columnwidth]{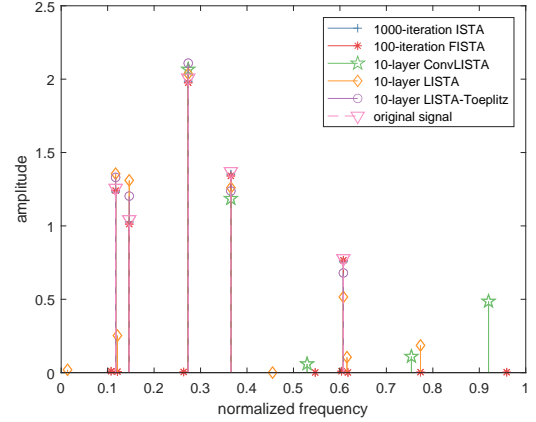}
	\caption{Recovered results of multiple harmonic components in an on-the-grid case via LISTA and 1D LISTA-Toeplitz network.}
	\label{fig:ongrid-1D}
\end{figure}

\begin{figure}
	\centering
	\includegraphics[width=0.9\columnwidth]{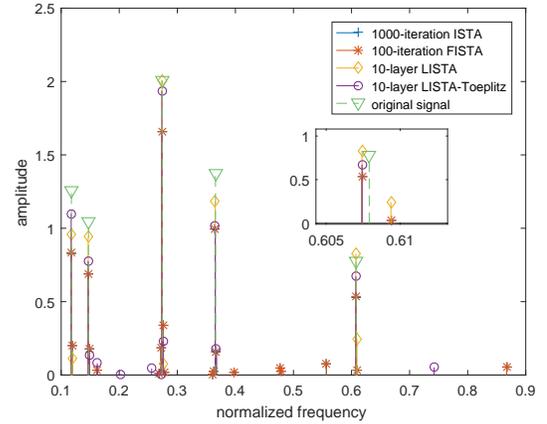}
	\caption{Recovered results of multiple harmonic components in an off-the-grid case via 1D LISTA-Toeplitz network.}
	\label{fig:offgrid-1D}
\end{figure}

Firstly, we consider noiseless cases, and examine the recovery performance using a single trial. The frequencies of true sinusoids are set on and off the predefined frequency points, respectively, with their recovery results shown in Figs.~\ref{fig:ongrid-1D} and \ref{fig:offgrid-1D}.

In our simulation, we set 5 targets with random amplitudes, marked by triangles.
In the on-the-grid case, ConvLISTA can only recover 2 targets of highest amplitude, while all the other methods (ISTA, FISTA, LISTA and LISTA-Toeplitz) can successfully recover all of them. It is because imposing the Toeplitz structure constraint on the linear block ${\bm W_e}\bm y$ leads to bad recovery performance in 1D harmonic retrieval problem.
We also find that LISTA-Toeplitz has close performance with other methods, successfully recovering the on-the-grid and off-the-grid sinusoids, while it has a significant reduction of network parameter number compared with the traditional \ac{lista}. 
\edit{We also present simulation results to show that our LISTA-Toeplitz is able to recover the off-the-grid frequencies.}
Comparing with Fig.~\ref{fig:ongrid-1D}, where frequencies are on-the-grid, we shift the true frequencies one quarter of the grid size away from the on-the-grid values in Fig.~\ref{fig:offgrid-1D}. Particularly, we set the actual frequencies by $f_{k} = (m_k+1/4)/M$, where $m_k$ is the grid index nearest to the $k$-th sinusoid, $k = 1,2,\dots,K$. As revealed by previous study \cite{Chi2011Mismatch}, the off-the-grid case causes mismatch between the assumed and the actual frequencies, which brings higher sidelobe pedestal problem. However, the proposed LISTA-Toeplitz method successfully recovers the neighboring grid points. 

\begin{figure} [tb]
	\centering
	\includegraphics[width=0.9\columnwidth]{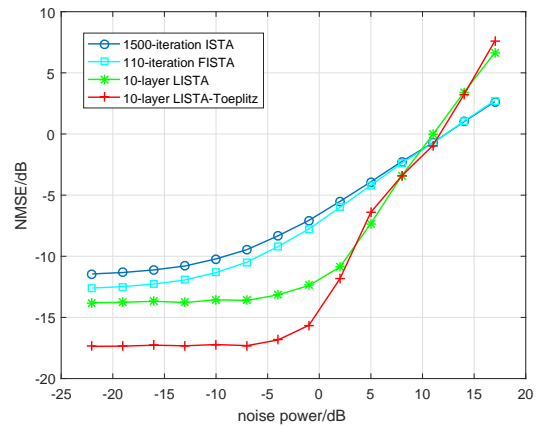}
	\caption{The NMSEs of 1D harmonic retrieval between four different methods (ISTA, FISTA, LISTA, LISTA-Toeplitz).}
	\label{fig:compareNMSE-1d}
\end{figure}

\begin{figure} [tb]
	\centering
	\includegraphics[width=0.9\columnwidth]{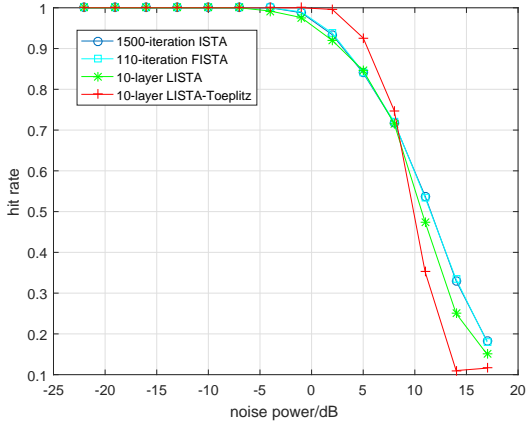}
	\caption{The hit rates of 1D harmonic retrieval between four different methods (ISTA, FISTA, LISTA, LISTA-Toeplitz).}
	\label{fig:comparehitrate}
\end{figure}

Secondly, we consider noisy cases, where we use \ac{nmse} and hit rate as performance metrics to evaluate the tested recovery methods versus noise power. 
Based on the same sparse signal with the previous experiment, we change the noise power, and calculate the \ac{nmse} and hit rates by averaging 1000 Monte Carlo trials, yielding Figs.~\ref{fig:compareNMSE-1d} and \ref{fig:comparehitrate}, respectively. 
As shown in Fig.~\ref{fig:compareNMSE-1d}, the NMSE of each method is decreasing monotonically with the noise power. When the noise power is no larger than 10 dB, the ten-layer LISTA and LISTA-Toeplitz achieve higher accuracy than that of the FISTA and ISTA, whose numbers of iterations are set 110 and 1500, respectively, much larger than the counterparts of learning-based methods. We also find that the proposed LISTA-Toeplitz has the lowest NMSE when the noise power is less than 0 dB, and has close performance with LISTA when the noise power is larger than 0 dB.
As shown in Fig.~\ref{fig:comparehitrate}, under reasonably small noise lever (less than 5 dB), the LISTA-Toeplitz network leads to highest hit rates, which shows its advantage on finding the frequencies of harmonics. 

\if false
\begin{figure} [tb]
	\centering
	\includegraphics[width=0.9\columnwidth]{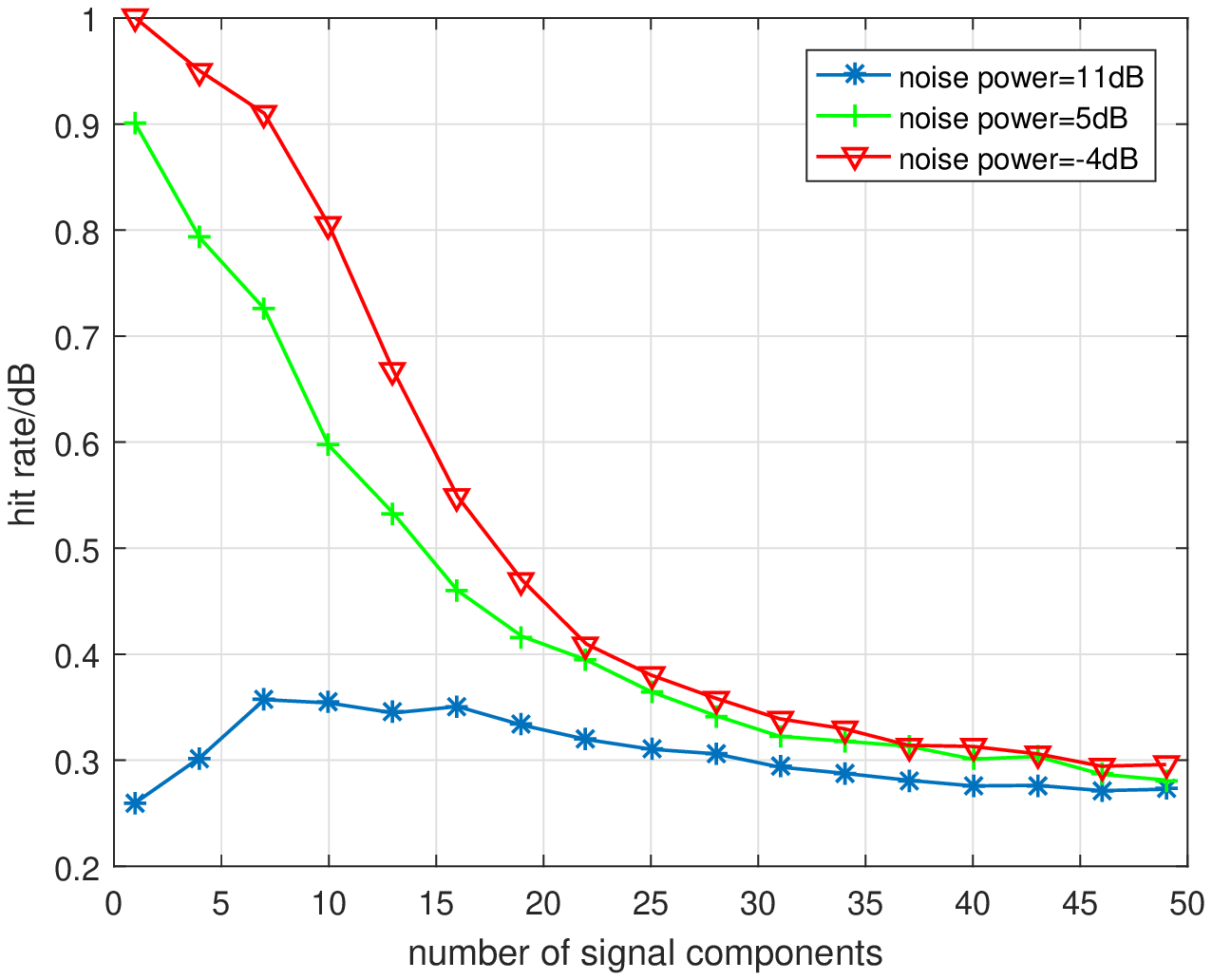}
	\caption{The hit rate versus the signal sparsity of 1D harmonic retrieval using LISTA-Toeplitz network.}
	\label{fig:hitrate_3line}
\end{figure}
\begin{figure} [tb]
	\centering
	\includegraphics[width=0.9\columnwidth]{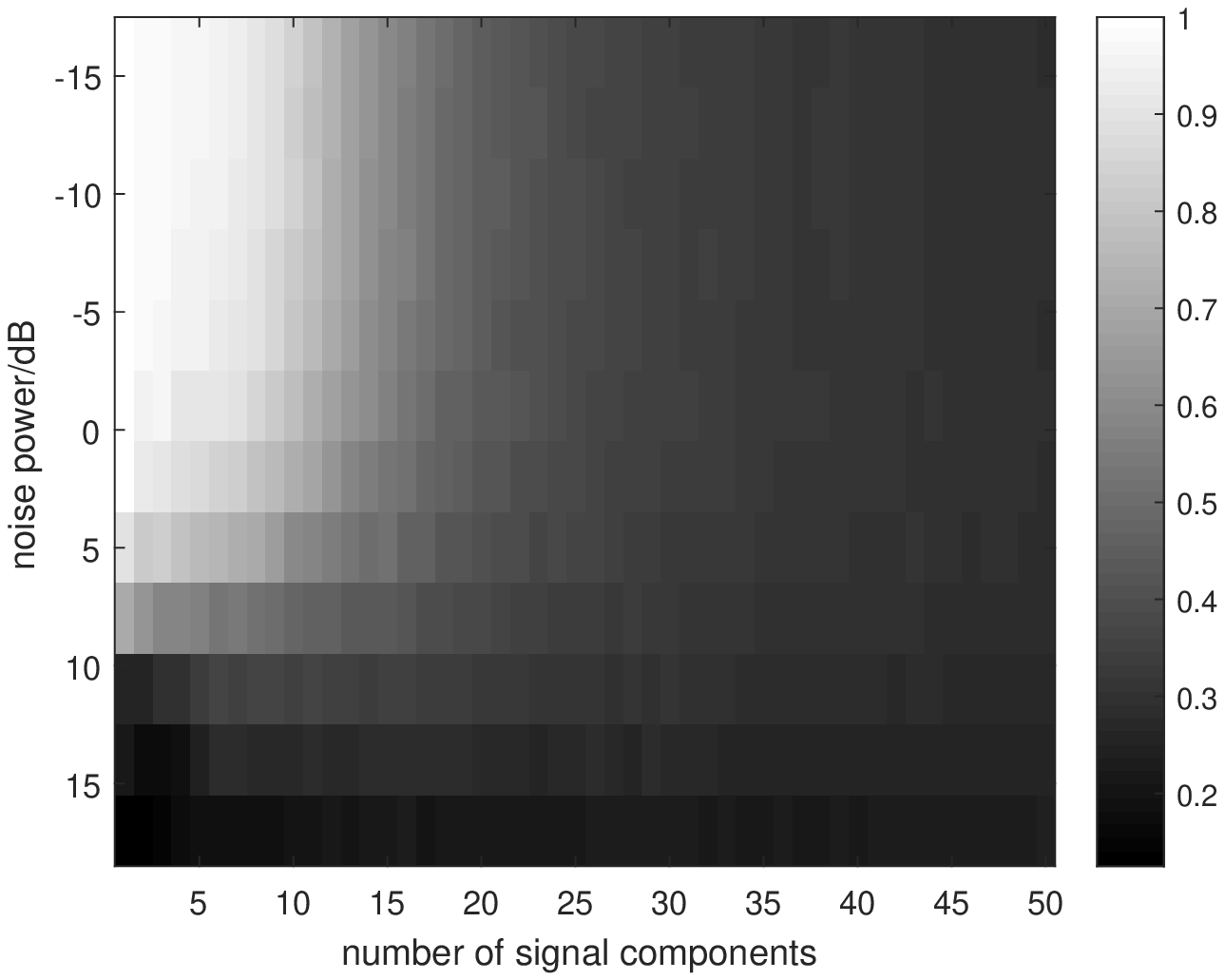}
	\caption{The hit rate of 1D harmonic retrieval using LISTA-Toeplitz network under different noise power and signal sparsity.}
	\label{fig:hitrate_map}
\end{figure}
\fi

\subsection{2D Harmonic Retrieval with Synthetic Data}\label{subsec:2Dsimulation}
{
	
	Simulations were also conducted to show the performance of LISTA-Toeplitz network for 2D harmonic retrieval problems.
	Similarly, we consider both noiseless and noisy cases. 
}

In the 2D case, simulations are configured as follows. The numbers of full observations in two dimensions are $M_1=8$ and $M_2=64$, respectively. From these overall $M_1 M_2$ observations, we randomly select $N=64$ samples. 
The number of distinct sinusoids is $K=5$. 
In these simulations, we generate 50,000 samples to train the ConvLISTA and our LISTA-Toeplitz network, while we use $10^7$ samples to train the ten-layer LISTA network. For both learned networks, we prepare 1000 samples for the validation and test sets.

\begin{figure}
	\centering
	\includegraphics[width=0.9\columnwidth]{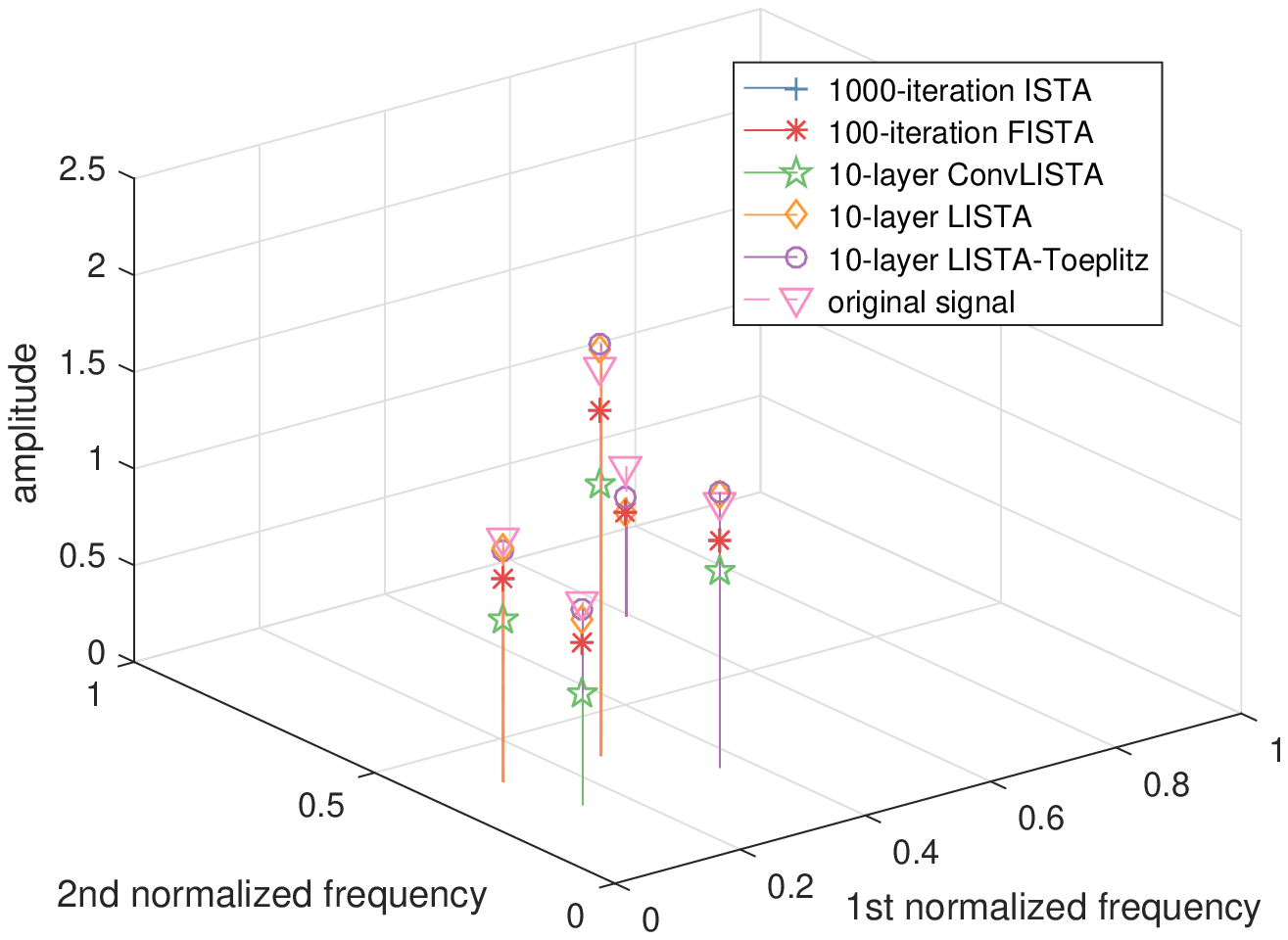}
	\caption{Recovered 2D plane of multiple harmonic components in an on-the-grid case via LISTA and 2D LISTA-Toeplitz network.}
	\label{fig:ongrid-2D}
\end{figure}

Similarly to the \ac{1d} case presented in Subsection \ref{subsec:1Dsimulation}, we first provide a single noiseless trial to intuitively demonstrate the recovery output of the tested algorithms. When the frequencies are all on the grid points, the results are shown in Fig.~\ref{fig:ongrid-2D}.
For clarity of the presentation, only components with amplitudes larger than 0.5 are shown in the plot, where we see that all the algorithms expect ConvLISTA work well in reconstructing sparse signals. 

In our simulation, we set 5 targets with random amplitudes, marked by triangles.
In Fig.~\ref{fig:ongrid-2D}, ConvLISTA can only recover 4 targets of highest amplitude, while all the other methods succeed in recovering all the 5 targets. Thus, to guarantee the recovery performance in harmonic retrieval problems, it is suggested to only make the block for the mutual inhibition part to be convolutional and remain the others as standard linear.

\begin{figure}
	\centering
	\includegraphics[width=0.9\columnwidth]{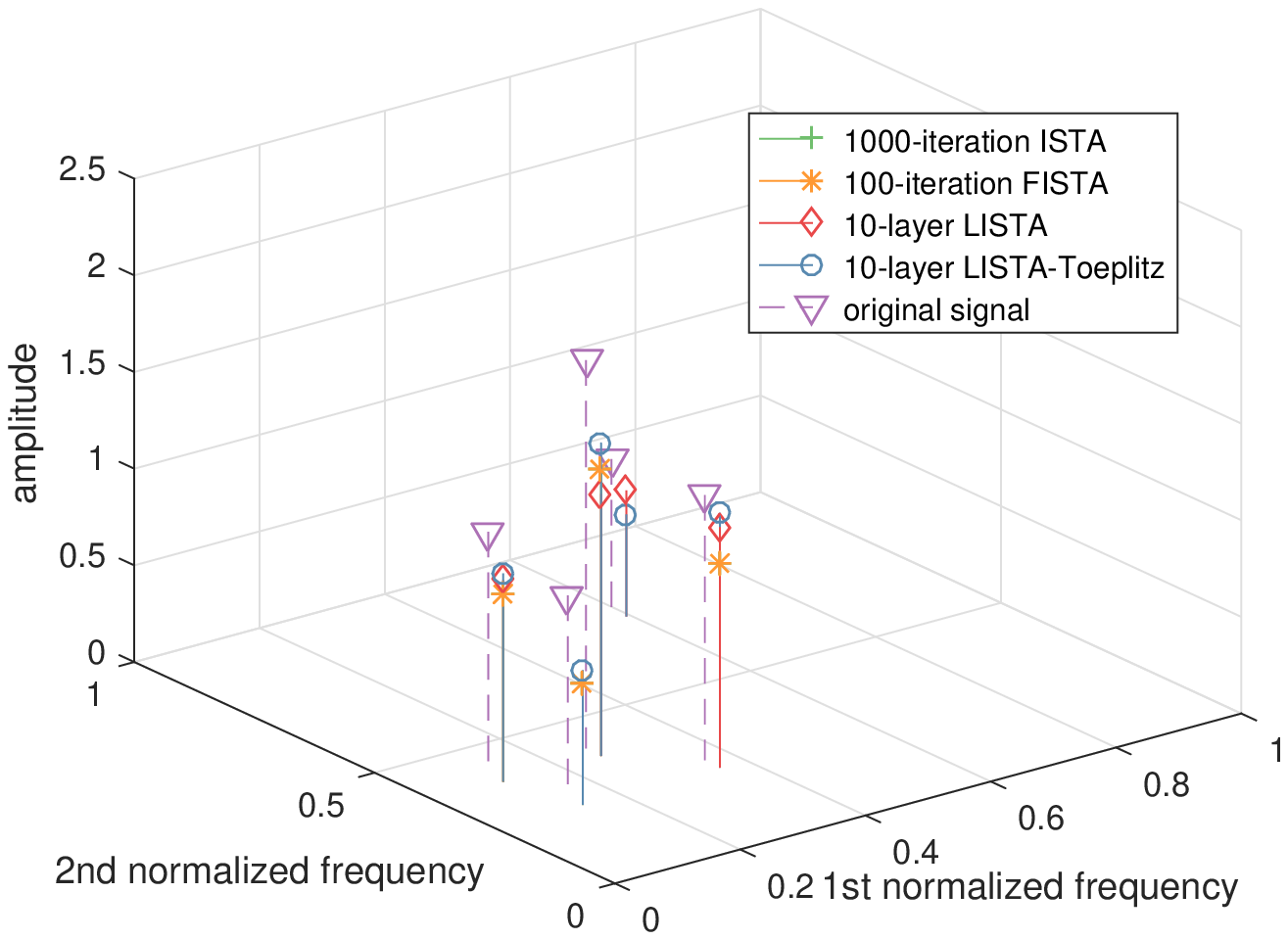}
	\caption{Recovered 2D plane of multiple harmonic components in an off-the-grid case via 2D LISTA-Toeplitz network.}
	\label{fig:offgrid_2d}
\end{figure}

We also consider the case when the frequencies are off the grid. Here, the frequency values are one quarter of the grid size away from its nearest grid point along the second frequency dimension, i.e., $f_{p,k} = (m_k+1/4)/M_p$, where $m_k$ is the grid index of the $k$-th sinusoid, $p =2$, $k = 1,2,\dots,K$. 
Fig.~\ref{fig:offgrid_2d} presents the results of an off-the-grid example,
and shows that this four methods successfully find the nearest grid points corresponding to those off-the-grid ground truth.
Particularly, LISTA-Toeplitz network achieves comparable performance with other methods by learning a smaller number of network parameters.

\begin{figure} [tb]
	\centering
	\includegraphics[width=0.9\columnwidth]{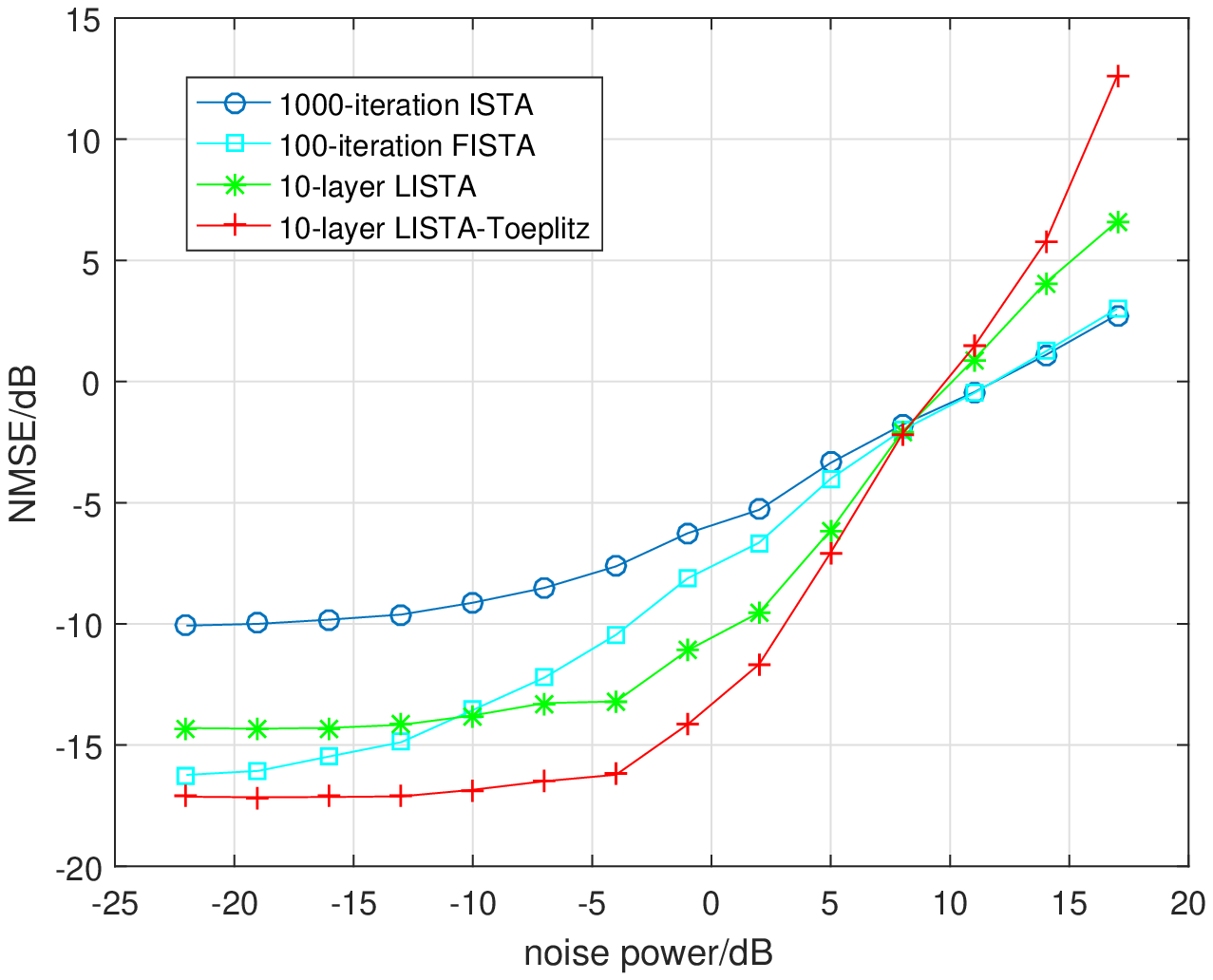}
	\caption{Simulation results of four methods (ISTA, FISTA, LISTA, LISTA-Toeplitz) on the 2D harmonic retrieval problem in terms of NMSE.}
	\label{fig:RVcompareNMSE}
\end{figure}

\begin{figure} [tb]
	\centering
	\includegraphics[width=0.9\columnwidth]{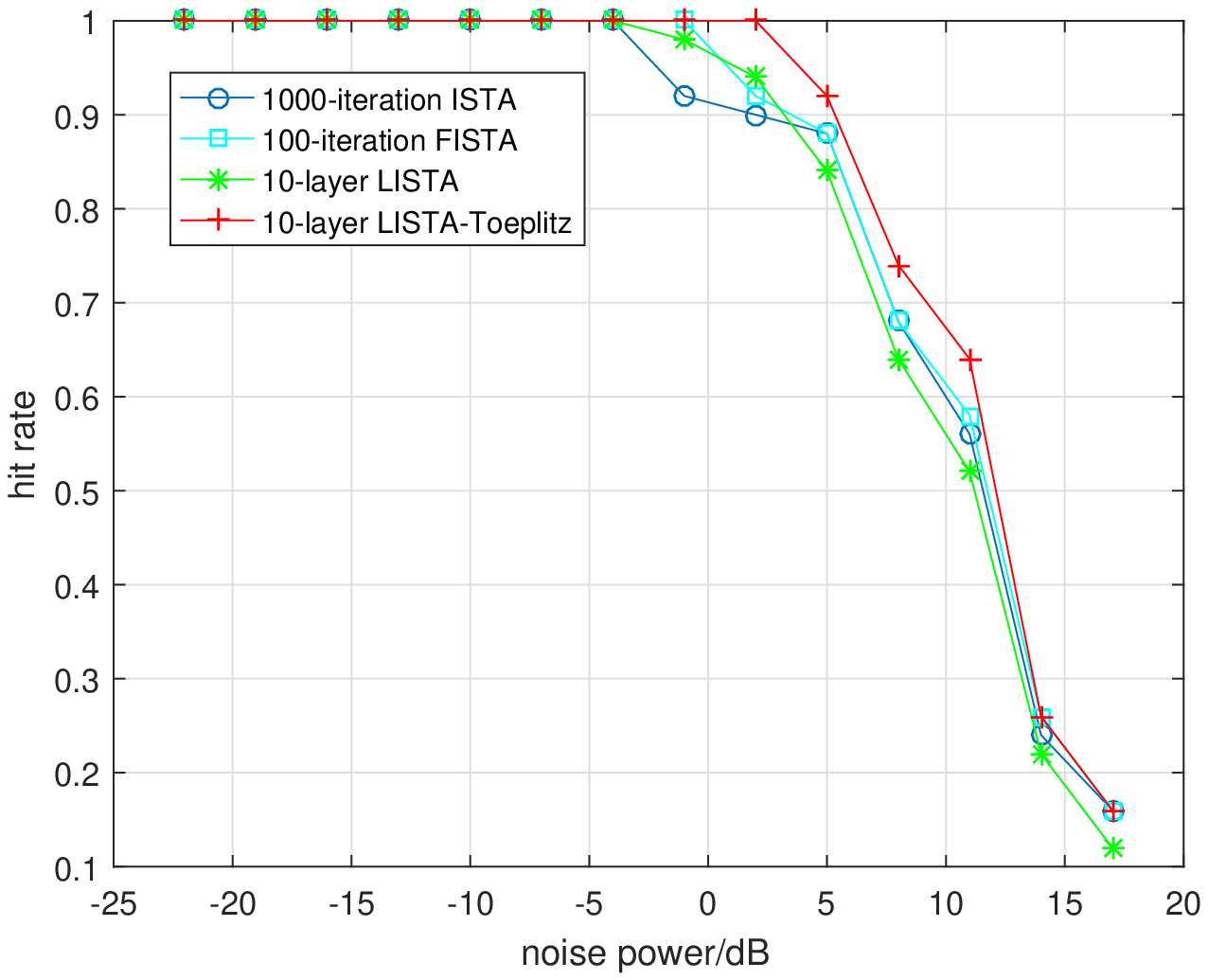}
	\caption{Simulation results of four methods (ISTA, FISTA, LISTA, LISTA-Toeplitz) on the 2D harmonic retrieval problem in terms of hit rate.}
	\label{fig:RVcomparehitrate}
\end{figure}

We then consider noisy scenarios, and evaluate the performance of tested methods versus noise power in terms of both NMSE and hit rates, leading to simulation results shown in Figs.~\ref{fig:RVcompareNMSE} and \ref{fig:RVcomparehitrate}, respectively.
{
	When the noise power is lower than $-5$ dB, the NMSE of the LISTA-Toeplitz network is $-17.2$ dB, lower than the other methods. 
	From Figs.~\ref{fig:RVcomparehitrate}, the hit rate curves of the four methods are close: The hit rates of ISTA and FISTA are almost the same, slightly lower than the LISTA-Toeplitz network.}
In comparison to original \ac{ista} and \ac{fista}, LISTA and LISTA-Toeplitz network improve upon ISTA by achieving a similar result using one to two orders of magnitude fewer iterations, while LISTA-Toeplitz improves upon LISTA with faster training and higher reconstruction quality.


\subsection{Experimental data set for Air Target Recovery}\label{subsec:real data}
{
	
	To further investigate the performance of the 2D LISTA-Toeplitz network, experimental scenarios were carried out, where we used \ac{rsfr} to detect civil aircraft in air and estimate the range-Doppler parameters of its dominant scatterers. The configuration of the radar is introduced in \cite{ExtendedTarget}. Here, we formulate the involved range-Doppler reconstruction problem as a \ac{2d} harmonic retrieval problem, and resort to the proposed 2D LISTA-Toeplitz network. The range and Doppler domain correspond to two frequency domains, with $M_1 = 64$ and $M_2 = 16$. The observations, the radar echoes of transmitted $N = 64$ pulses, are regarded samples of the virtually full $M_1 M_2$ observations \cite{HuangLXEW18}.
	
	Under such settings, we compare the 2D LISTA-Toeplitz network with the conventional ISTA method, disregarding the standard LISTA network, because the LISTA network has too many variables (more than $10^7$ parameters) to learn and it is hard to train such a huge LISTA network.
}





\begin{figure}
	\centering 
	\subfigure[]{ 
		\includegraphics[width=0.9\columnwidth]{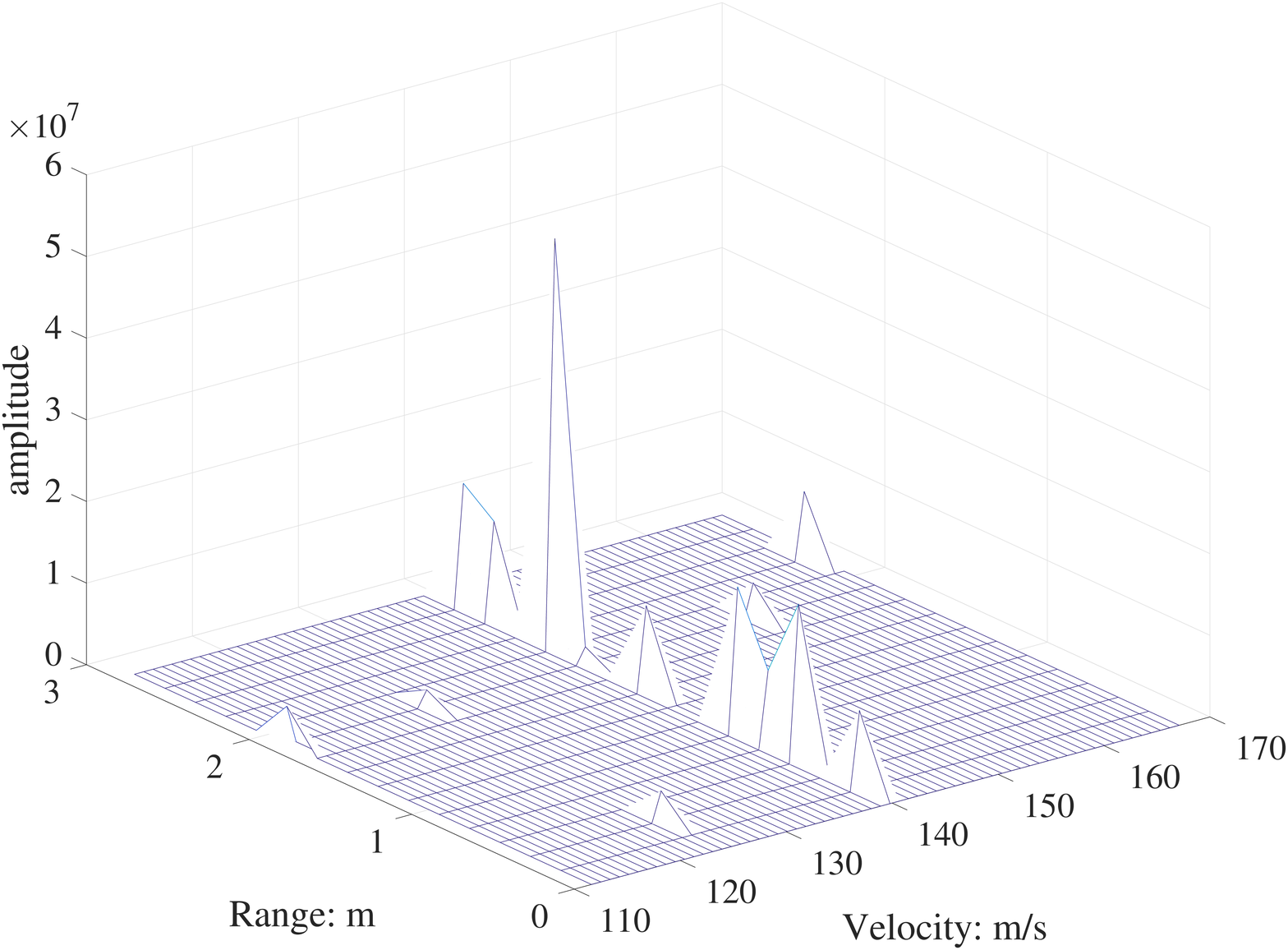}
	} 
	\subfigure[]{ 
		\includegraphics[width=0.9\columnwidth]{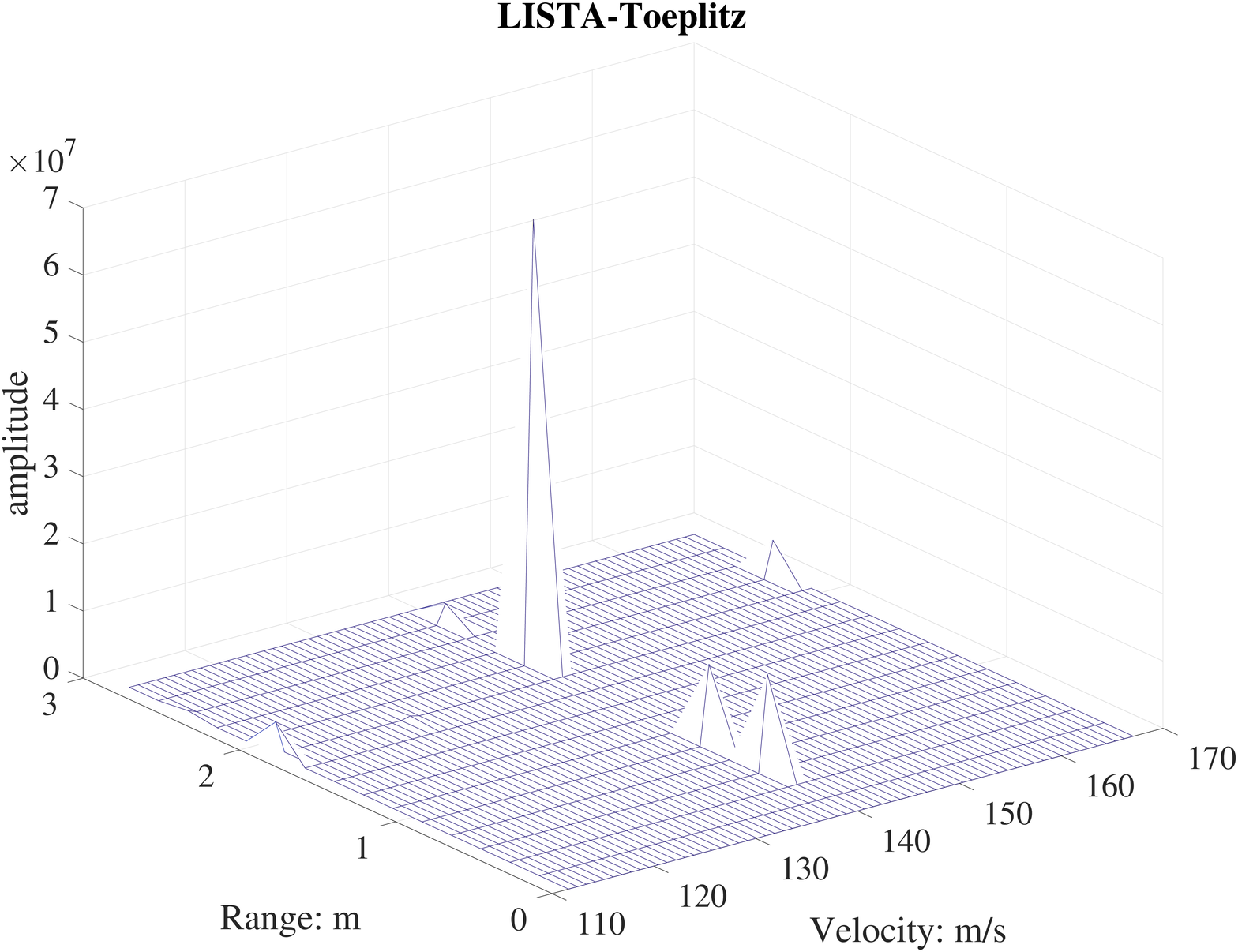}
	} 
	\caption{The reconstructed air target using (a) ISTA and (b) LISTA-Toeplitz.}
	\label{fig:aircraft}
\end{figure}

The recovered range-Doppler parameters of scatterers are indicated in Fig.~\ref{fig:aircraft}. Both ISTA and LISTA-Toeplitz recover some scatterers laid on the grids of velocity $140$~m/s, which accords with the fact that the aircraft is flying away from the radar with a radial relative velocity approximately of $140$ m/s. 
This experiment validates the correctness of the proposed method in real applications.

\if false
\subsection{Discussions}\label{subsec:advantages}
\Verified{I suggest you omit this part. Because it is quite difficult to understand these methods and the disadvantages. If they are applicable and important in the applications of harmonic retrieval, reviewers may argue that you should numerically compare the performance with these methods. Instead, you may briefly mention some of most relevant methods in the introduction section.
	\\
	Regarding the comparison between LISTA and LISTA Toeplitz in terms of space and time complexity, you may move the content corresponding section.
}
{
	After showing the validness of our proposed structured network, here we put emphasis on some advantages of our proposed structured network over conventional algorithms and other deep learning networks.
}


Compared to conventional iterative algorithms (e.g. ISTA or FISTA), data-driven networks can operate when not aware of the exact dictionary. 
In practical application, although sometimes we can have some knowledge of the dictionary, there may be some errors in the expected dictionary, which can lead to high model error. 
Unlike conventional algorithms, LISTA and LISTA-Toeplitz 
can be ignorant of the knowledge of the model.
They have the ability to extract meaningful model information from the training data \cite{BengioAI} so that they are certifiably robust against model uncertainty.



Compared with other types of network compression approaches, our structured network compression shows several unique advantages. 

Firstly, unlike weight-pruning methods\cite{Wang2019NonstructuredDW,AnwarStructuredPrun}, our structured network based on the model structure is very competitive in terms of both storage and computation efficiency. 
Unstructured weight-pruning methods (i.e. simply constrain a fraction of elements of network's weights to remain zero during training) produce irregular pruned networks and incurs index accesses due to its irregular network connections \cite{Wang2019NonstructuredDW}.
Moreover, even when the sparsity of a network's weights is well structured, it also requires extra re-training to compensate for the losses due to pruning, which brings additional complexity to the training process \cite{AnwarStructuredPrun}. 
However, our structured construction approach guarantees the strong structure of the trained network, thereby avoiding the computation time overhead incurred by the complicated indexing or re-training process.

\emph{Discussion:}
Here we introduce some observations about limitations of the deep-learning method when solving \ac{mhr} problems.
\begin{enumerate}
	\item{Basis mismatch} 
	
	The fundamental assumption in compressive sensing is the sparsity of the underlying signal in the basis of representation. 
	However, a mismatch between the assumed and the actual basis may cause the signal to appear as incompressible. One such example is the mismatch of angle grids in DOA estimation due to the inadequate discretization of the DOA domain. When the sources do not coincide with the points on the selected angular grid (particularly in the case of moving sources), the signal might not appear sparse in the selected angle grid due to spatial spectral leakage. 
	
	Since the fundamental assumption of sparsity is violated, sparse reconstruction algorithms might have poor performance under basis mismatch. 
	An analysis of the sensitivity of compressive sensing to basis mismatch is found in \cite{Chi2011Mismatch}. 
	\item{Coherence and network performance} 
	
	Herein, we assume that the compressive sensing problem is discretized densely enough to avoid basis mismatch and we study its limitations due to a coherent basis.
	Two well-known measurements of correlation between any two columns of the dictionary matrix are its mutual coherence 
	and the restricted isometry property (RIP)\cite{CAHILL2016363} which is described by the restricted isometry constants.
	
	As a dictionary-learning method, the performance of LISTA is related to RIP property of the dictionary. The more ill-conditioned dictionary matrix we use, the worse recovery result the LISTA network will finally learn. 
	For example, in DOA estimation problem, as the columns or equivalently the rows of the Gram matrix of the dictionary matrix represent the beam-pattern for the corresponding focusing DOA direction, just have a look of two Gram matrices with the same size for both bad and good cases as below (just changing the grid step).
	
	\begin{figure}[H]
		\centering 
		\subfigure[]{ 
			\includegraphics[width=1.6in]{figure/ill-conditioned_dict.eps}
		} 
		\subfigure[]{ 
			\includegraphics[width=1.6in]{figure/good-conditioned_dict.eps}
		} 
		\caption{the Gram matrix of the dictionary matrix for (a) bad case with angular grid step of $1^\circ$, and (b) good case with angular grid step of $30^\circ$.}
		\label{fig:output}
	\end{figure}
	So LISTA and LISTA-Toeplitz cannot handle compressive sensing problems with strong coherence in dictionaries.
	A simple way to achieve better recovery performance is to decrease the coherence, for example, in DOA estimation which means to use a sparse array with a large aperture.
\end{enumerate}

\fi

\section{Conclusion}\label{sec:conclusion}
This paper considered the compressive \ac{mhr} problem, which estimates frequency components in multidimensional spectra based on a compressive measurement.
We proposed a structured network to solve \ac{mhr} problems by revealing and exploiting the Toeplitz structure in the Gram matrix of the measurement matrix. Compared with the traditional LISTA network, the proposed network, namely LISTA-Toeplitz network, greatly reduces the dimension of network variables, which consequently reduces the time and space complexity and makes the network easier to train.
Other convolutional extensions such as ConvLISTA which is not designed based on the specific signal model will suffer major degradation on recovery performance.
We presented both simulated and real data of 1D/2D harmonic retrieval problems to verify the performance of the proposed algorithms.
In the numerical examples, our network shows excellent recovery performance in terms of both NMSE and hit rate, better than ConvLISTA and comparable to ISTA, FISTA and LISTA.


\appendix

\section{Network Training Details} \label{sec:train}

In this appendix, we illustrate some additional specifications when constructing and training a \ac{lista}-Toeplitz network for complex-value applications like harmonic retrieval. 
In Appendix~\ref{subsec:complexextension}, we give a detailed description of the necessary extension for our proposed network when applied to complex-value cases.
{
	Some details in training process is provided in Appendix \ref{subsec:training}, including dataset generation, parameter initialization and some training strategies.
	
}

\subsection{Necessary Extensions for Proposed Structured Network} \label{subsec:complexextension}
{
	
	In Section \ref{sec:design}, we redesign the original LISTA network by converting multiplications of Toeplitz matrices to linear convolutions. 
	Here we further extend it to a complex-value framework, because most off-the-shelf deep learning toolboxes, e.g., PyTorch and TensorFlow, are only applicable to real-value networks. 
	Particularly, we transform the complex-value linear convolution as well as matrix multiplication and non-linear operations to their real-value counterparts. 
}

\if false
Here we just take the 1D LISTA-Toeplitz network with two layers as an example, where the first layer computes a linear transformation followed by a nonlinear soft-threshold operator and the second layer is an output layer with $\ell2$ loss. For simplicity, we ignore the bias computation ${{\bm W_e}}\bm y$ in \eqref{eq:LISTA} and only concentrate on the gradient computation of $\bm W_g$.

When training the LISTA network, 
the partial gradient of the error function $J(\bm x)$ with respect to each entry of ${\bm W_g}$ is
\begin{eqnarray} 
\frac{{\partial J\left( \bm x \right)}}{{\partial {W_g}\left( {i,j} \right)}} = \left( {{\text{S}_\theta }\left( {{\bm W_g}\left( {i,\cdot} \right)\bm x} \right) - x_i^{\mathrm{true}}} \right){x_j}, i,j = 1,2,\cdots,M.
\end{eqnarray}
Here we ignore the derivative of the soft-threshold operator, which is a step function $Step(x) = 1$ when $\left| x \right| \ge \theta$.
As the gradient with respect to networks with multi-layers can be simply computed with the chain rule (i.e. the well-known Back-Propagation algorithm), It suffices therefore to just analyze the gradient-based optimization for the two-layer network shown above.

For the LISTA network, we need to compute the gradient under the constraint ${\bm W_g(i,j)} = h_{i-j}$ so that all the connections which share the same weight should be summed up. This gives us
\begin{eqnarray} 
\begin{aligned}
\frac{{\partial J\left(\bm x \right)}}{{\partial {h_i}}} &= {\left( {\bm x \gg i} \right)^T}\left( {{\text{S}_\theta }\left( {{\bm W_g}\bm x} \right) - \bm x^{\mathrm{true}}} \right), \\
i &= -(M-1),\cdots, -1,0,1,\cdots,(M-1)
\end{aligned}
\end{eqnarray}
where 
${\bm x \gg i}$ means shifts the vector by $i$ element in the direction of right if $i<0$ (i.e. downwards for a column vector) or left if $i>0$ (i.e. upwards for a column vector) while having zeros padded as necessary.

Therefore, using the flip-and-filter interpretation of linear convolution, we can have
\begin{eqnarray} \label{eq:gradient}
{\nabla _{\bm h}}J\left(\bm x \right) = flip(\bm x) * \left( {{\text{S}_\theta }\left( {{\bm W_g}\bm x} \right) - \bm x^{\mathrm{true}}} \right)
\end{eqnarray}
which means that the partial gradient w.r.t. structured weights correspond to filtering the input vector $x$ with the gradient w.r.t. the layer's output.
\fi

{
	Different from some existing complex-value network extensions \cite{Complex1,Complex2}, which separately feed a double-size real-valued network with the real and imaginary parts of the inputs, discarding the links between real and imaginary parts, 
	we maintain the structure of the complex-value data and operations, as discussed in the sequel. 
}
\if false
The main difficulty of transformation from the complex domain to the real domain lies in the part of the non-linear operator while the linear operators (for example, complex matrix multiplication and linear convolution operator) are in the same manner as their real-value operators.

As for complex matrix multiplication, it holds that
\begin{eqnarray}	
\begin{aligned}
\bm W\bm x & = ({\bm W_\mathrm{R}} + \mathrm{j}{\bm W_\mathrm{I}})({\bm x_\mathrm{R}} + \mathrm{j}{\bm x_\mathrm{I}})\\
& = ({\bm W_\mathrm{R}}{\bm x_\mathrm{R}} - {\bm W_\mathrm{I}}{\bm x_\mathrm{I}}) + \mathrm{j}({\bm W_\mathrm{R}}{\bm x_\mathrm{I}} + {\bm W_\mathrm{I}}{\bm x_\mathrm{R}}),
\end{aligned}
\end{eqnarray}
where ${({\cdot})_\mathrm{R}}$ and ${({\cdot})_\mathrm{I}}$ denote the real and imaginary parts of complex data, respectively. 

As for complex linear convolution operator, it holds that
\begin{eqnarray}	
\begin{aligned}
\bm h*\bm x & = ({\bm h_\mathrm{R}} + \mathrm{j}{\bm h_\mathrm{I}})*({\bm x_\mathrm{R}} + \mathrm{j}{\bm x_\mathrm{I}})\\
& = ({\bm h_\mathrm{R}}*{\bm x_\mathrm{R}} - {\bm h_\mathrm{I}}*{\bm x_\mathrm{I}}) + \mathrm{j}({\bm h_\mathrm{R}}*{\bm x_\mathrm{I}} + {\bm h_\mathrm{I}}*{\bm x_\mathrm{R}}).
\end{aligned}
\end{eqnarray}

Denoting $\left| \bm x \right|$ as $f(\bm x_\mathrm{R},\bm x_\mathrm{I})$, it error gradients holds that
\begin{eqnarray}
\begin{array}{l}
\begin{aligned}
\frac{{\partial f\left( {{\bm x_\mathrm{R}},{\bm x_\mathrm{I}}} \right)}}{{\partial {\bm x_\mathrm{R}}}} &= \frac{{{\bm x_\mathrm{R}}}}{{f\left( {{\bm x_\mathrm{R}},{\bm x_\mathrm{I}}} \right)}},\\
\frac{{\partial f\left( {{\bm x_\mathrm{R}},{\bm x_\mathrm{I}}} \right)}}{{\partial {\bm x_\mathrm{I}}}} &= \frac{{{\bm x_\mathrm{I}}}}{{f\left( {{\bm x_\mathrm{R}},{\bm x_\mathrm{I}}} \right)}},
\end{aligned}
\end{array}
\end{eqnarray}
\fi

The main difficulty is on the non-linear operator of the complex numbers in each layer, i.e., the soft-threshold operator defined in \eqref{eq:Soperator}, which can be rewritten as
\begin{equation}\label{eq:Soperator2}
\mathcal{S}_{\theta}\left( [\bm{x}]_i \right) 
= [\bm{x}]_i \left(1 - \frac{\theta}{\max \left( \left| [\bm{x}]_i \right|,\theta \right)} \right),
\end{equation}
which equals $0$ when $\left| [\bm{x}]_i \right| < \theta$, and yields $\mathcal{S}_{\theta}\left( [\bm{x}]_i \right) 
	= [\bm{x}]_i \left(1 - \frac{\theta}{\left| [\bm{x}]_i \right|} \right) = {\rm sign}([\bm{x}]_i)(\left| [\bm{x}]_i \right| - \theta)$ when $\left| [\bm{x}]_i \right| \ge \theta$. 
To be compatible with the real-value deep learning toolboxes, we realize \eqref{eq:Soperator2} with real-value components. To this end, we first compute the absolute value of $\bm x$. 
Then, the real and imaginary parts of \eqref{eq:Soperator2} are computed with
\begin{eqnarray}
\begin{array}{l}
\begin{aligned}
\Re\{\mathcal{S}_{\theta}\left( [\bm{x}]_i \right) \} &= \Re\{[\bm{x}]_i\} { \left(1 - \frac{ \theta}{\max \left( \left| [\bm{x}]_i\right|, \theta \right)} \right)},\\
\Im\{\mathcal{S}_{\theta}\left( [\bm{x}]_i \right) \} &= \Im\{[\bm{x}]_i\} { \left(1 - \frac{ \theta}{\max \left( \left| [\bm{x}]_i \right|, \theta \right)} \right)},
\end{aligned}
\end{array}
\end{eqnarray}
respectively, where $\Re\{\cdot\}$ and $\Im\{\cdot\}$ denote the real and imaginary parts of a complex argument, respectively. 

For the linear matrix multiplication and convolution operator, we implement their complex-value counterpart operators through two cross coupling real-value channels, e.g., for \ac{2d} linear convolution operator,
$
\bm H *\bm X = (\Re\{\bm H\}*\Re\{\bm X\} - \Im\{\bm H\}*\Im\{\bm X\}) + \mathrm{j}(\Re\{\bm H\}*\Im\{\bm X\} + \Im\{\bm H\}*\Re\{\bm X\})
$. 

\begin{figure} [tb]
	\centering
	\includegraphics[width=0.9\columnwidth]{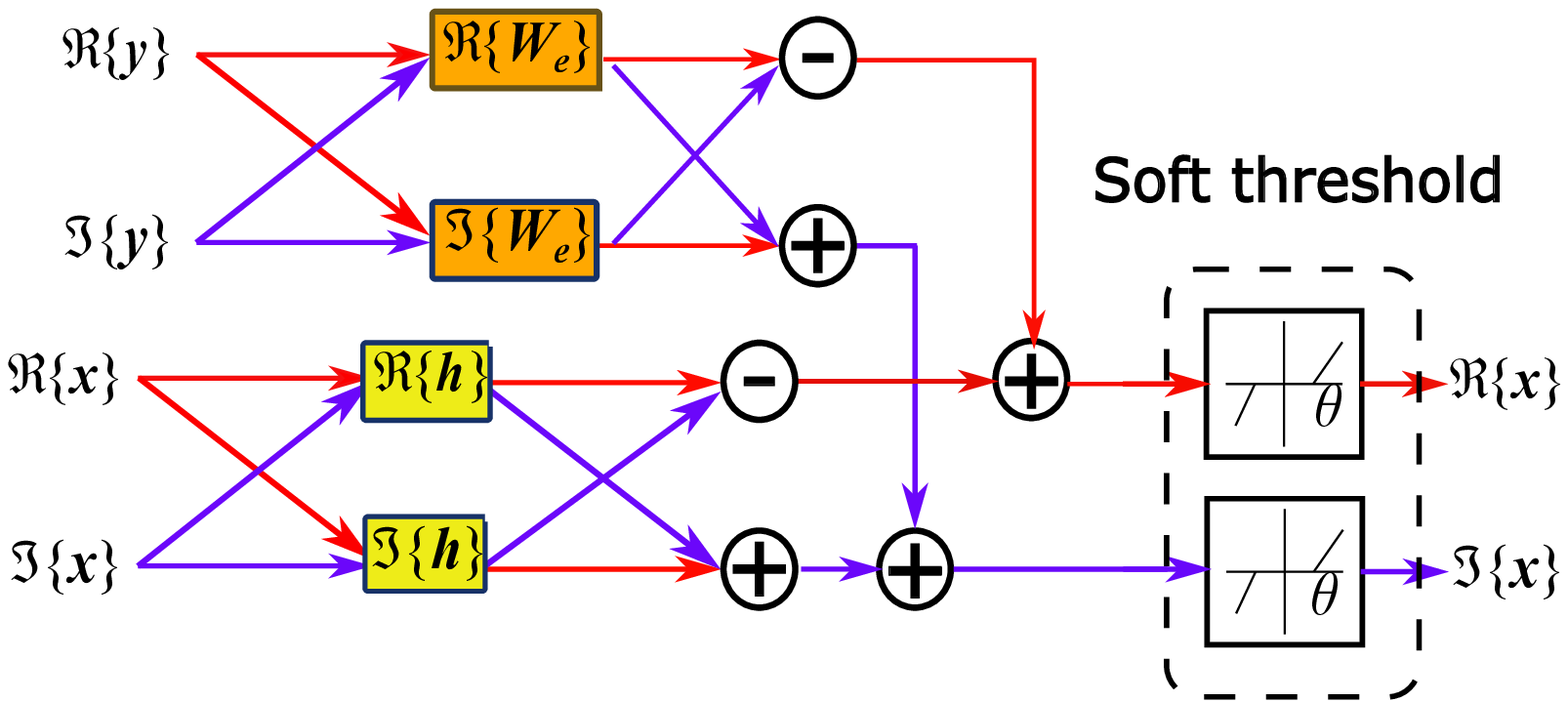}
	\vspace{-0.31cm}
	\caption{Block diagrams of one layer of the LISTA-Toeplitz network applied to complex-value data.}
	\label{fig:complexframe}
\end{figure}

{
	With above preparations, we construct each layer of the complex-value LISTA-Toeplitz network with real-value components, as illustrated in Fig.~\ref{fig:complexframe}, which facilitates the use of off-the-shelf framework toolboxes. 
}

\subsection{Training Details} \label{subsec:training}
{
	
	In this appendix, we describe some main details when training our proposed network, including three main parts: 1) generate data set, 2) initialize network parameters, and 3) training strategies, as will be introduced in the sequel.
	
}

\subsubsection{Data Preparation} \label{subsec:Data}
	To evaluate the performance of the proposed method, we generate three data sets: training, validation and test sets. 
We first prepare $N_{tr}$ training samples with known labels, i.e., the ground truth of sparse signal $\bm x$. In these training samples, we denote by $\left\{ \bm y_i \right\}_{i = 1}^{N_{tr}}$ and $\left\{ \bm x_i \right\}_{i = 1}^{N_{tr}}$ the observation and label sets, respectively, which obey $\bm y_i = \bm \Phi \bm x_i$, $i = 1,2,\cdots,N_{tr}$. 
Denote two matrices $\bm Y \in \mathbb{C}^{N \times N_{tr}}$ and $\bm X \in \mathbb{C}^{M \times N_{tr}}$ by
\begin{eqnarray}
\begin{array}{l}
\begin{aligned}
\bm Y &= {\left[ \bm{y}_1 \quad \bm{y}_2 \quad \cdots \quad \bm{y}_{N_{tr}} \right]},\\
\bm X &= {\left[ \bm{x}_1 \quad \bm{x}_2 \quad \cdots \quad \bm{x}_{N_{tr}} \right]}.
\end{aligned}
\end{array}
\end{eqnarray}
The choice of $N_{tr}$ is related to the initialization of network parameters, and will be discussed in the subsequent subsection.

After training process is finished, we generate another $N_{vl}$ and $N_{ts}$ samples for validation and testing, respectively. The validation data set is used for determining some hyper-parameters of the network. 

\subsubsection{Network Parameter Initialization} \label{subsec:Init}
{
	
	It is important to initialize network parameters correctly, because a good starting point allows the network to converge rapidly from the very beginning. 
	In a LISTA-Toeplitz network, each layer has a set of parameters to learn, e.g., $\{ {\bm W_e}, {\bm h} ,{\theta ^{(k)}}\} $ for a \ac{1d} LISTA-Toeplitz network or $\{ {\bm W_e}, {\bm H} ,{\theta ^{(t)}}\} $ for a 2D counterpart.
}

Following the steps in \cite{OnsagerLAMP}, we initialize the network parameter ${\bm W_e} \in {\mathbb{C}^{M \times N}}$ by generating a coarse estimate of the dictionary matrix from training pairs, which is computed as
\begin{eqnarray}
\widehat {\bm \Phi} = \bm Y (\bm X^H\bm X)^{-1}\bm X^H.
\end{eqnarray}
Then, the learned matrix ${\bm W_e}$ is initialed as ${\bm W_e} =\frac{1}{L} \widehat {\bm \Phi} ^T$, where $L = {\lambda _{\max }}( {{\widehat {\bm \Phi} ^H}\widehat {\bm \Phi} } )$. 

The initial values of ${\bm W_g}$ (for LISTA), ${\bm h}$ or ${\bm H}$ (for \ac{1d} or \ac{2d} LISTA-Toeplitz, respectively) are chosen as zeros in the experiments. They can also be initialized by random values.

The initialization also affects the required number of training data. 
It has been widely accepted as a practical criterion in many published prediction modeling studies that the amount of samples need to be more than roughly ten times the degrees of freedom in the model \cite{ZhuVFR15,VanSampleSize,lei2019how}. 
For a $T$-layer \ac{1d} LISTA-Toeplitz network, the numbers of parameters in $\bm W_g$ and $\bm W_e$ are around $O(2MT)$ and $O(MNT)$, respectively. Here, since $\bm W_e$ has a good choice of initial value, it can be learnt efficiently with a small amount of training samples. 
However, $\bm W_g$ is initialed as zeros or random numbers, which requires to be update from training data. 
Therefore, in our simulation we generate around $N_{tr} = O(20MT)$ pairs of sparse signals and corresponding measurement signals to obtain a well-trained network.

\subsubsection{Training Protocol} \label{subsec:Procedure}

With these $N_{tr}$ labeled training data and initialized parameters, we train the network implemented using TensorFlow with the strategy described below.

We choose \ac{nmse} of training samples as a quantitative metric, also referred to as the loss function, which is defined as 
\begin{eqnarray}\label{eq:NMSE}
	\mathrm{NMSE} = 
	{\sum\limits_{i = 1}^{N_{tr}} {\left\| {\bm x_i - \bm{\widehat x}_i} \right\|_2} }/
	{\sum\limits_{i = 1}^{N_{tr}} {\left\| \bm x_i \right\|_2} },
\end{eqnarray}
where $\bm{\widehat x}_i$ is the output signal reconstructed by the proposed neural network with respect to the observation $\bm y_i$.

Then we start optimizing the network parameters towards minimizing the loss function on the training data set using Adam optimizer.

\bibliographystyle{IEEEtran}
\bibliography{IEEEabrv}

\end{document}